*Article*

# Fracture Detection in Wrist X-ray Images Using Deep Learning-Based Object Detection Models

**Fırat Hardalaç** [1], **Fatih Uysal** [1,2,]*, **Ozan Peker** [1], **Murat Çiçeklidağ** [3], **Tolga Tolunay** [3], **Nil Tokgöz** [4], **Uğurhan Kutbay** [1], **Boran Demirciler** [5] **and Fatih Mert** [5]

[1] Department of Electrical and Electronics Engineering, Faculty of Engineering, Gazi University, TR 06570 Ankara, Turkey; firat@gazi.edu.tr (F.H.); ozan.peker@gazi.edu.tr (O.P.); ukutbay@gazi.edu.tr (U.K.)

[2] Department of Electrical and Electronics Engineering, Faculty of Engineering and Architecture, Kafkas University, TR 36100 Kars, Turkey

[3] Department of Orthopaedics and Traumatology, Faculty of Medicine, Gazi University, TR 06570 Ankara, Turkey; muratciceklidag@gazi.edu.tr (M.Ç.); tolgatolunay@gazi.edu.tr (T.T.)

[4] Department of Radiology, Faculty of Medicine, Gazi University, TR 06570 Ankara, Turkey; nil.tokgoz@gazi.edu.tr

[5] Huawei Turkey R&D Center, TR 34768 İstanbul, Turkey; boran.demirciler@huawei.com (B.D.); fatih.mert@huawei.com (F.M.)

* Correspondence: uysal@gazi.edu.tr; Tel.: +90-534-022-6128

**Abstract:** Hospitals, especially their emergency services, receive a high number of wrist fracture cases. For correct diagnosis and proper treatment of these, images obtained from various medical equipment must be viewed by physicians, along with the patient's medical records and physical examination. The aim of this study is to perform fracture detection by use of deep-learning on wrist X-ray images to support physicians in the diagnosis of these fractures, particularly in the emergency services. Using SABL, RegNet, RetinaNet, PAA, Libra R-CNN, FSAF, Faster R-CNN, Dynamic R-CNN and DCN deep-learning-based object detection models with various backbones, 20 different fracture detection procedures were performed on Gazi University Hospital's dataset of wrist X-ray images. To further improve these procedures, five different ensemble models were developed and then used to reform an ensemble model to develop a unique detection model, 'wrist fracture detection-combo (WFD-C).' From 26 different models for fracture detection, the highest detection result obtained was 0.8639 average precision (AP50) in the WFD-C model. Huawei Turkey R&D Center supports this study within the scope of the ongoing cooperation project coded 071813 between Gazi University, Huawei and Medskor.

## 1. Introduction

Examination of the parts where bone fractures occur reveal that there are many cases of fractures on various parts of the body, such as wrists, shoulders and arms, especially in the emergency services of hospitals. It is also observed that the fractures that may result from various reasons and can occur as partial or complete fractures of the bones. These are classified as open or closed bone fractures. In an open fracture, also called a compound fracture, the skin is exposed through a deep wound, or the bone pierces the skin, becoming visible. In a closed fracture, also called a simple fracture, the bone is broken, with the skin remaining intact. It was concluded upon examination of the causes of bone fractures that fractures mostly occur when more force is applied to the bone than it can withstand. Based thereon, the causes that may lead to bone fractures are falls, trauma or a direct blow or kick to the body. Stress fractures that are common in athletes are caused by overuse or repetitive motions making the muscles tired, applying more pressure on the bone.

Moreover, fractures can also be caused by diseases that render the bone weak, such as osteoporosis or cancer in the bones. The symptoms of the bone fractures may include sudden pain, bruises, swelling, obvious deformity, warmth or redness [1].

Within the scope of the steps used in the detection of bone fractures, along with a complete medical history (including inquiry about the manner of occurrence of the fracture) and physical exam, physicians may require tests used for fractures. Mainly, three different devices are used for these tests, which are X-ray, MRI (Magnetic Resonance Imaging) and CT (Computed Tomography) [1]. The most preferred among these devices is the X-ray device, which is also more cost-efficient compared to the other options. X-ray images obtained from Gazi University Hospital were used for deep-learning based fracture detection in wrist images in this study.

The anatomy of the wrist consists of radius, ulna and carpal bones. The carpal bones connect the hand to the forearm. Figure 1 shows a sample image of the anatomy of the wrist.

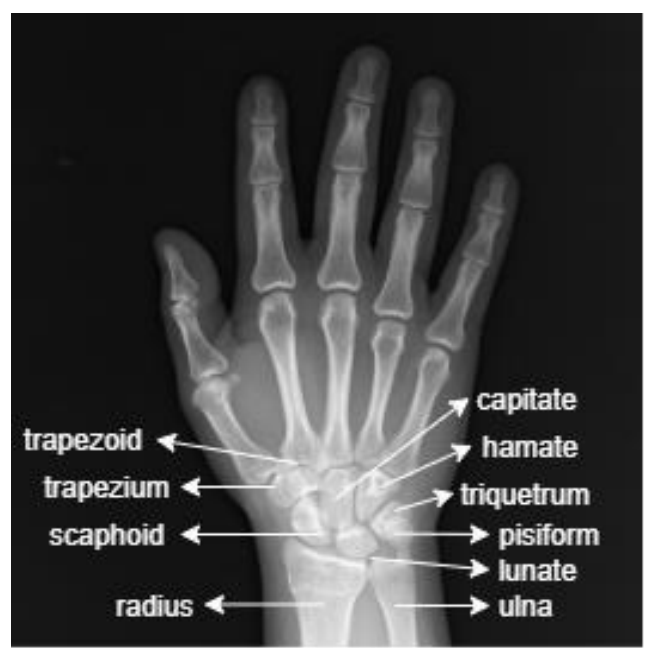

**Figure 1.** The anatomy of the wrist.

The image in Figure 1 shows that the wrist is made up of eight small carpal bones in total. These are: scaphoid, lunate, trapezium, trapezoid, capitate, hamate, triquetrum and pisiform.

- Scaphoid: A long, boat-shaped bone under the thumb.
- Lunate: A crescent-shaped bone beside the scaphoid.
- Trapezium: A rounded, square-shaped bone above the scaphoid and under the thumb.
- Trapezoid: The bone beside the trapezium shaped like a wedge.
- Capitate: An oval bone in the middle of the wrist.
- Hamate: The bone under the pinky finger side of the hand.
- Triquetrum: The pyramid-shaped bone under the hamate.
- Pisiform: A small, round bone that sits on top of the triquetrum.

Under these carpal bones are located the radius and ulna bones [2]. The fractures in the wrist X-ray images used in this study are focused on the radius and ulna bones.

The main contributions of this study, in which fracture detection is performed on wrist X-ray images using deep-learning-based object detection models, are as follows:

- In this study, fracture detection was performed in the first place, using 10 different deep learning models available in the literature, and the results thereof were compared.
- Many different augmentation methods were tried with these models, and the most compatible augmentation method was identified, on X-ray images in particular, followed by fracture detection procedures again.
- In order to better analyze the results of fracture detection, in addition to the average precision (AP) and average recall (AR) parameters that are widely used in detection procedures in the literature, localization recall precision (LRP), which is another up-to-date evaluation parameter, was used in the field of biomedical and in evaluation on bone fracture detection for the first time.
- Ensemble models were used to improve the results of detection procedures carried out with a total of 20 models with/without augmentation.
- Outputs of 10 different models with augmentation were evaluated from five different aspects in the development of ensemble models which are: AP, AR, single stage, two stage and LRP.
- With the new ensemble model developed within the scope of this study, an approximately 10% increase was achieved compared to the best AP result of the currently available detection models in the literature.
- A clinical dataset (non-public) obtained from Gazi University Hospital was used in the study, which improves the applicability of the study in hospitals.
- The augmentation method used in this study and/or an ensemble model with a logic similar to the ensemble model developed within the scope of this study can be used for similar detection studies that may be conducted in the future on X-ray images, particularly in the field of biomedical research.

## 2. Related Works

The literature includes fracture detection studies based on artificial intelligence on both open source and clinical bone image datasets collected from various medical devices. The average precision (AP) score obtained from the fracture detection performed by Guan et al., using a dilated convolutional feature pyramid network (DCFPN) on 3842 thigh fracture X-ray images is 82.1%. [3]. The highest result of fracture detection performed by Guan et al., on approximately 4000 arm fracture X-ray images in a musculoskeletal radiograph (MURA) dataset, 62.04% AP, was obtained using proposed two-stage region-based convolutional neural networks (R-CNN) method [4]. The AP50 score achieved by Wang et al., was 87.8% with the ParallelNet method developed for fracture detection in a dataset of 3842 thigh fracture X-ray images, using a TripleNet backbone network [5]. Using a part of the dataset of 1052 bone images in total, Ma and Luo carried out fracture detection with Faster R-CNN, followed by fracture/non-fracture classification with the proposed CrackNet model using the entire dataset, achieving an accuracy of 90.11% [6]. The AP score achieved as the result of the fracture detection performed by Wu et al., on a dataset consisting of 9040 hand, wrist, pelvic, knee, ankle, foot, shoulder and ankle radiographs in total by Feature Ambiguity Mitigate Operator (FAMO) model used with ResNeXt101 and feature pyramid network (FPN) was 77.4% [7]. Qi et al., carried out fracture detection procedures, achieving a mean AP (mAP) score of 68.8%, using the anchor-based Faster R-CNN model with multi-resolution FPN and ResNet50 backbone network for a total of 2333 femoral fracture X-ray images with nine different types of fracture [8]. Thian et al., used Inception-ResNet Faster R-CNN model for fracture detection on 7356 wrist radiographic images [9]. Sha et al., achieved a mAP score of 75.3% using the You Only Look Once (YOLOv2)-based model developed for fracture detection in a dataset of 5134 spinal fracture CT images [10]. With the fracture detection performed using another model based on Faster R-CNN developed by Sha et al., for the same dataset, the mAP achieved was 73.3% [11]. The sensitivity achieved as the result of segmentation and detection performed with the proposed U-Net-based FracNet in a total of 7473 rib fracture CT images collected from 900 patients by Jin et al., was 92.9% [12]. The AP score with the proposed guided anchoring method Faster R-CNN model for fracture detection in 3067 hand fracture X-ray images performed by Xue et al., was 70.7% [13].

In addition to studies on fracture detection, there are a number of studies on deep-learning-based classification in the literature in which the class of fracture is identified. Uysal et al., performed 26 different deep-learning-based classification procedures to determine the class of fracture in shoulder bone X-ray images in the musculoskeletal radiograph (MURA) dataset, and then developed two different ensemble learning models to further improve the results of the classification [14]. Raghavendra et al., achieved a classification accuracy of 99.1% with the proposed CNN model in the thoracolumbar CT dataset containing 420 normal images and 700 fracture images [15]. Within the scope of fracture classification performed by Beyaz et al., on 1341 femoral neck X-ray images using the proposed CNN model and genetic algorithm (GA), the accuracy achieved was 79.3% [16]. Tobler et al., carried out fracture classification with the ResNet18 model in the dataset consisting of 15,775 frontal and lateral radiographs, resulting in the highest accuracy of 94% [17]. As the result of classification performed to identify the class of fracture in 1389 wrist radiographs using the InceptionV3 model, Kim et al., achieved an area under the receiver operator characteristic curve (AUC) score of 0.954 [18]. Chen et al., achieved an accuracy of 73.59% for the vertebral fracture class in a dataset of 1306 plain frontal radiographs using the ResNeXt model [19]. Within the scope of the fracture classification performed by Tanzi et al., in 2453 proximal femur X-ray images using InceptionV3, VGG16 and ResNet50 models, the highest accuracies achieved for structures of grade three and grade five were 87% and 78%, respectively [20]. Öksüz et al., proposed a segmentation network in which training is carried out that optimizes three different tasks in cardiac MR images: image artefact detection, artefact correction and image segmentation [21]. A distance for structural similarity metric and Fuzzy C-Means algorithm were developed for image segmentation by Tang et al. [22].

It is concluded based on the studies available in the literature that the classification models such as ResNet and VGG used to identify the class of fracture are used as backbone networks in deep-learning-based object detection models. Moreover, it is observed upon examination of the studies in the literature on detection of bone fractures that mostly Faster R-CNN or YOLO-based detection including backbone networks, such as ResNet or ResNeXt, is performed on CT and X-ray images. In this study, fracture detection procedures are performed on wrist X-ray images collected from Gazi University Hospital. For detection, Deformable Convolutional Networks (DCN), Dynamic R-CNN, Faster R-CNN, Feature Selective Anchor-Free (FSAF), Libra R-CNN, Probabilistic Anchor Assignment (PAA), RetinaNet, RegNet and Side-Aware Boundary Localization (SABL) models with various backbone networks were used in the first place. Based on the results of bone detection obtained therein, ensemble models were developed for fracture detection in wrist X-ray images for better detection results, providing a contribution to the literature.

The third part of the study explains the deep-learning-based object detection models and the proposed ensemble object detection models used for fracture detection. The fourth part of the study, titled "Experiments," includes the dataset collected from Gazi University Hospital used within the scope of the study and labelling of the site of fracture, as well as data preprocessing and data augmentation, AP, AR, LRP scores obtained using the fracture detection models and predicted bounding boxes and precision-recall curves. The fifth and final part of the study, titled "Conclusion and Future works," explains the contribution of the ensemble model developed within the scope of the study to the literature, as well as potential future developments.

## 3. Methods

Within the scope of the study, deep-learning-based object detection models consisting of various backbone networks were used and developed for the detection of fracture areas in wrist X-ray images. Firstly, deep-learning-based DCN (Faster R-CNN), Dynamic R-CNN, Faster R-CNN, FSAF, Libra R-CNN (RetinaNet), PAA, RetinaNet, RegNet (RetinaNet) and SABL (Faster R-CNN and RetinaNet) models available in the literature, as well as RegNet RegNetX-3.2GF, as backbones for these models, and ResNet50 was used for object detection for the models. Therefore, a total of 20 different procedures of fracture detection were performed with and without data augmentation with these models.

The transfer learning method was implemented in all deep-learning-based models used for fracture detection. With the use of the transfer learning method, the current weight of each detection model pre-trained with the COCO dataset was used. COCO is a dataset containing a significant number of images and various object categories, in which object detection and segmentation can be performed [23]. In the deep learning models used for detection in Figure 2, the number of object categories was reduced from 80 to 1. The reason for this is while there are 80 object categories in the models trained with the COCO dataset, there is only 1 object category in this study as the fracture detection is performed in the wrist X-ray images.

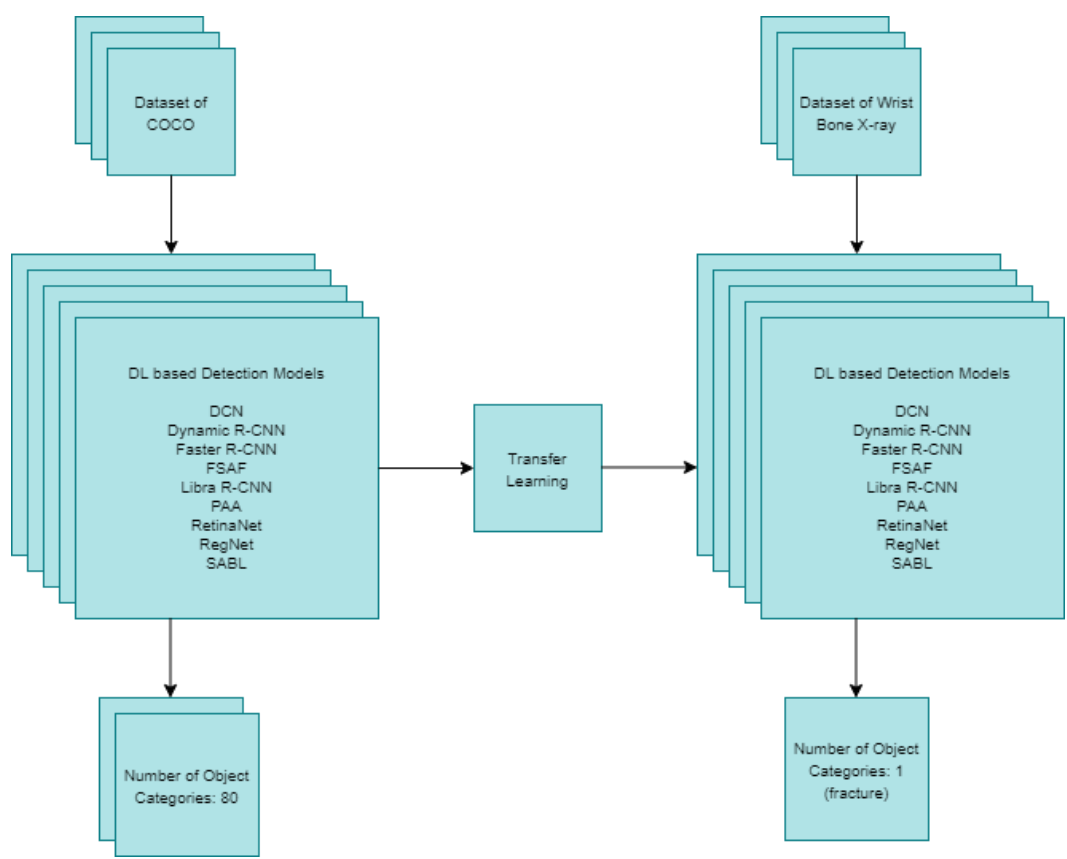

**Figure 2.** Deep-learning based detection models using transfer learning.

Based on an examination of the results obtained using the detection procedures in Figure 2, new ensemble models were developed to further improve the results of the fracture detection. The details of deep-learning-based built object detection models used for fracture detection in wrist X-ray images and the ensemble models developed are provided in sub-headings as follows.

*3.1. Fracture Detection Models Based on DL for Wrist X-ray Images*

Firstly, deep-learning-based built object detection models were used for fracture detection in wrist X-ray images. The models used at this stage have different structures, but are mainly single-stage and two-stage. The general structure of the structures used within the scope of the study is provided in Figures 3 and 4 below.

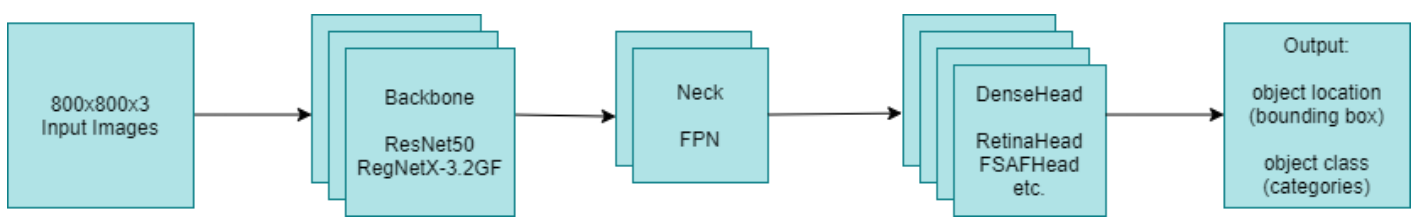

**Figure 3.** Single-stage object detectors (RetinaNet, FSAF, etc.).

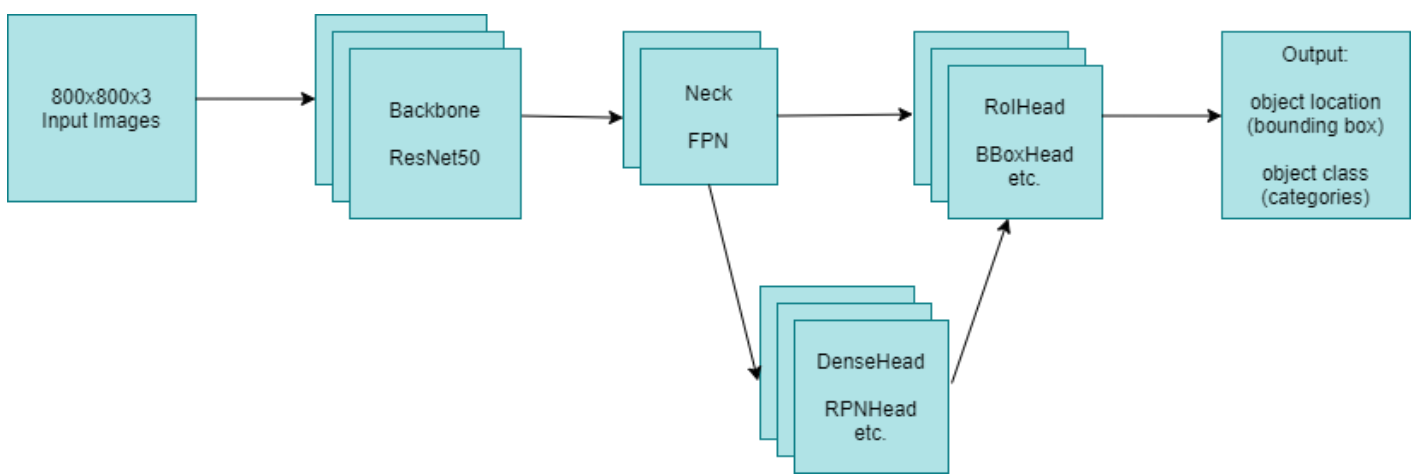

**Figure 4.** Two-stage object detectors (Dynamic R-CNN, Faster R-CNN, etc.).

When the general structure is shown in Figure 3 regarding single-stage object detectors is examined, it is seen that the object location and object class can be obtained at the output by first passing the input images through the backbone and then through the neck and DenseHead operations, respectively. When looking at the two-stage object detectors structure specified in Figure 4, although there are similar structures such as backbone and neck in single-stage object detectors, the most basic difference between the two is that they also contain RoIHead in addition to the backbone, neck and DenseHead in two-stage detectors. In addition, when examined in terms of training times, in general, it is understood that single-stage detectors can detect faster than two-stage detectors. When examined in terms of accuracy, although the training time of two-stage detectors is slower, they generally have higher accuracy than single-stage detectors. However, this may vary depending on the dataset.

In both models' stages, it is seen that FPN is used in the neck parts and ResNet50 is used in the backbone. Additionally, RegNetX-3.2GF can be used in single-stage object detectors. FPN is a feature extractor that is used in object detection operations and is independent of backbone convolutional architectures. In FPNs, they generate proportionally sized feature maps at multiple levels [24]. RegNetX is used as a backbone network in RegNet, one of the object detection models. RegNetX is a convolutional network design space where there is a linear parameterization of block widths [25]. In this study, RegNetX-3.2GF was used in the RegNet model used for fracture detection. ResNet, on the other hand, is a deep learning model that contains more than one residual block [26]. The number of these blocks may vary depending on the number of layers. In the object detection models used in this study, ResNet50 with 50 layers was used in all models where ResNet was used as the backbone network.

Regarding the sample outputs of both types of object detection models, the bbox numbers produced by the single-stage PAA and two-stage Dynamic R-CNN models at certain threshold values in the [0.1, 0.9] range are given in Table 1. In addition, more detailed bbox numbers for these two models are available in Figure 5. When the results are examined in detail, PAA produces a low probability value but many bboxes, while the Dynamic RCNN model produces a high probability value but few outputs. This is clearly understood when looking at the test results from 0.1 to 0.9 of the threshold value.

**Table 1.** PAA and Dynamic R-CNN bbox numbers according to different threshold values.

| Threshold Value | PAA Bbox | Dynamic R-CNN Bbox |
| --- | --- | --- |
| 0.1 | 4317 | 291 |
| 0.2 | 1098 | 176 |
| 0.3 | 415 | 124 |
| 0.4 | 159 | 97 |
| 0.5 | 75 | 79 |
| 0.6 | 36 | 71 |
| 0.7 | 4 | 65 |
| 0.8 | 0 | 57 |
| 0.9 | 0 | 48 |

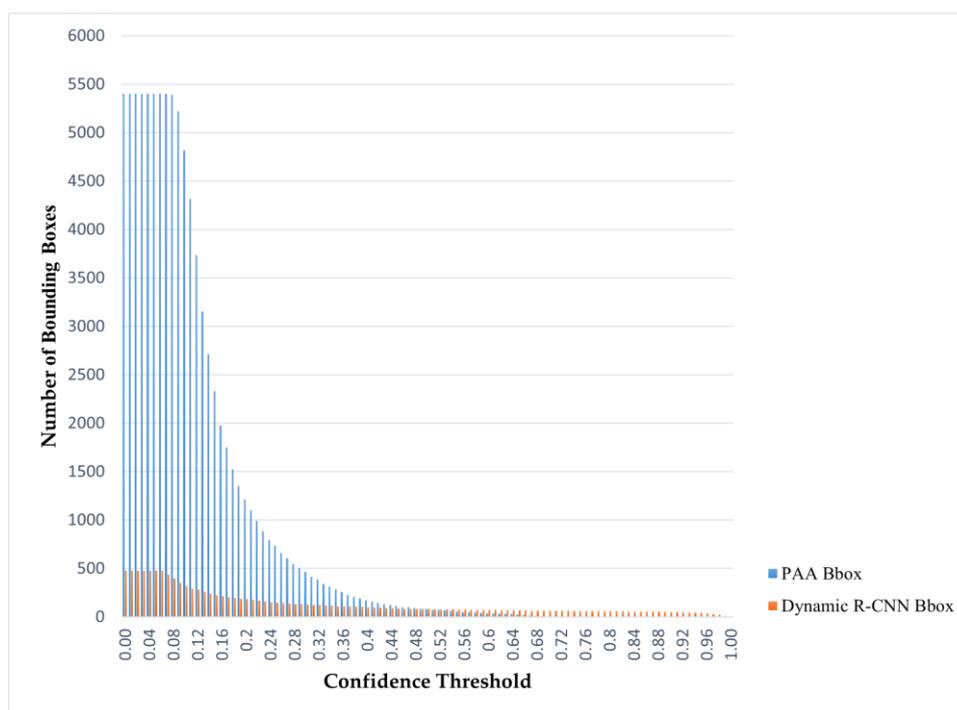

**Figure 5.** PAA and Dynamic R-CNN bbox numbers according to different threshold values.

FSAF, PAA, RetinaNet and RetinaNet-based models (SABL, Libra, RegNet) were used as single-stage models, and Dynamic R-CNN, Faster R-CNN and Faster R-CNN-based models (SABL, DCN) were used as two-stage models. The subheadings below provide details of the models used within the scope of this study.

3.1.1. RetinaNet

RetinaNet is a single-stage deep-learning-based object detection model that can achieve higher detection results compared to many different two-stage object detection models. RetinaNet models mainly consist of ResNet in the backbone, Feature Pyramid Network (FPN) in the neck and RetinaHead in the DenseHead, that is, the classification subnet and box regression subnet [27]. In this study, a RetinaNet object detection model pretrained with the COCO dataset with a ResNet50 backbone was used for fracture detection.

### 3.1.2. Feature Selective Anchor-Free (FSAF)

Feature Selective Anchor-Free (FSAF) is an anchor-free model for single-shot object detectors, consisting of a simple and effective building block. It is concluded upon examination of FSAF models that they are RetinaNet based. FSAF uses ResNet as backbone, FPN in neck and FSAFHead in the bounding box head [28]. Different from RetinaNet, this study uses FSAFHead as DenseHead in FSAF.

### 3.1.3. Dynamic R-CNN

Dynamic R-CNN consists of two main components, the Dynamic Label Assignment (DLA) process and the Dynamic SmoothL1 Loss (DSL). The IOU threshold is increased automatically in a dynamic manner to improve the quality of proposals in the DLA. In DSL, the shape of the regression loss function is adjusted to ensure such improvement [29]. In the Dynamic R-CNN model used within the scope of this study, fracture detection was performed by pretraining with the COCO dataset with a ResNet50 backbone.

### 3.1.4. Libra R-CNN

Libra R-CNN consists of three main components to reduce any imbalance at sample, feature and objective levels, and to further improve detection performances. These components are IoU-balanced sampling, a balanced feature pyramid and balanced L1 loss. This model appears to improve detection results when applied in both Faster R-CNN and RetinaNet models [30]. The hardware used within the scope of this study allowed the use of the Libra RetinaNet model for fracture detection.

### 3.1.5. Faster R-CNN

R-CNN and Fast R-CNN use a protracted selective search affecting the network performance to find region proposals. In Faster R-CNN, a Region Proposal Network (RPN) is proposed to find the region proposals. This enables the development of a two stage detector that can perform faster object detection compared to Fast R-CNN [31]. A Faster R-CNN two-stage detector with a ResNet50 backbone was used for fracture detection in this study.

### 3.1.6. Side-Aware Boundary Localization (SABL)

SABL is an approach that can be applied to both single-stage and two-stage detectors, which can localize each side of the bounding box with a dedicated network branch, respectively. This approach can be applied to Faster R-CNN, RetinaNet and Cascade R-CNN [32]. For detection procedures in this study, SABL was applied to both Faster R-CNN and RetinaNet with a ResNet50 backbone.

### 3.1.7. Deformable Convolutional Networks (DCN)

DCN mainly consists of two modules, which are deformable convolution and deformable RoI pooling. Using these modules, the transformation modeling capacity of CNNs is improved. The procedures performed in object detection and instance segmentation suggest that this model has a positive contribution to the results [33]. DCNv2, that is, the 2nd version of this model, with Faster R-CNN and a ResNet50 backbone, was used for fracture detection within the scope of this study.

### 3.1.8. Probabilistic Anchor Assignment (PAA)

With PAA, a new anchor assignment that can separate anchors into positive and negative samples with adaptable anchors, was proposed and applied. In this model, anchor scores are calculated in the first place, and the probability distributions of these scores are fitted. Subsequently, the model is trained with anchors that are separated into positive and negative samples based on their probabilities. Basically, in the PAA model, only a single convolutional layer is added to the RetinaNet model [34]. Within the scope of this study, fracture detection was performed using the version of the PAA model with a ResNet50 backbone.

3.1.9. RegNet

RegNet proposes and uses a network design paradigm where stage widths and depths are determined by the quantized linear function. With this paradigm, design spaces that combine the advantages of manual design and neural architecture search and parameterize network populations are designed. The RegNet model can be used with Faster R-CNN, Mask R-CNN and RetinaNet [25]. The hardware used in this study only allowed the use of the RegNet RetinaNet model with a RegNetX-3.2GF backbone as a RegNet model for fracture detection.

*3.2. Proposed Fracture Detection Models Based on DL for Wrist X-ray Images*

Based on the results of the detection of deep-learning-based models performed on wrist X-ray images and used for fracture detection within the scope of the study, ensemble models were used in order to further improve the results of fracture detection. While determining the submodels to be used for the ensemble model, 5 different cases were taken into consideration, consisting of optimal threshold values of AP50, AR and LRP, as well as the single-stage and two-stage structures of the models. Details of the ensemble model used and proposed specifically for the study are provided below.

Wrist Fracture Detection (WFD) Ensemble Model Based on Weighted Boxes Fusion

Firstly, five different Weighted Boxes Fusion-based WFD ensemble models were used. While the ensemble models developed and proposed using the submodels determined based on the single-stage and two-stage structures examined are WFD-1 and WFD-2, respectively, the ensemble models developed and proposed using the submodels determined based on the AP50, AR and LRP optimal threshold values are WFD-3, WFD-4 and WFD-5, respectively. While the WFD-3 ensemble model was developed using the detection models with the 5 best AP50 scores in total, consisting of Dynamic R-CNN, FSAF, Libra RetinaNet, PAA and RetinaNet, the WFD-4 ensemble model was developed with the models with the 5 best AR scores, consisting of FSAF, Libra RetinaNet, PAA, RetinaNet and SABL RetinaNet, using the most compatible models and weight coefficients determined out of 4715 different combinations with varying weight coefficients between 1–5. WFD-1 and WFD-2 ensemble models were developed, respectively, based on the most compatible models and weight coefficients determined out of 40,593 different combinations using weight coefficients between 1–6 based on 6 single-stage models and 388 different combinations using weight coefficients between 1–4 based on 4 two-stage models. WFD-5 ensemble models was developed based on the most compatible models and weight coefficients determined out of 4715 different combinations using weight coefficients between 1–5 using DCN, Faster R-CNN, FSAF, Libra RetinaNet and RetinaNet with the best LRP optimal threshold values. All WFD ensemble models developed are based on weight boxes fusion.

For the ensemble models named WFD-1,2,3,4,5, different trials were carried out in each WFD to determine the most performance model combinations. The steps followed while performing these operations is basically as follows: Step 1: First of all, the object detectors associated with the relevant WFD model and the number of them are determined. Step 2: Since the number of models required for the formation of the WFD model is at least two, different combinations are tried, from two to the maximum number of models. Step 3: The weights used during these combinations are for a maximum of n object detectors; this changes as an integer in the range of 1-n. Therefore, while determining the weight coefficients, it starts from 1 and increases consecutively by the maximum number of object detector models. For example, detectors associated with WFD-1 are single-stage models with RegNet, FSAF, RetinaNet, SABL RetinaNet, PAA and Libra RetinaNet, and their number is 6. Therefore, while the number of models is minimally 2 and maximally 6 in combinations made with WFD-1, the weight coefficients vary in the range of 1–6. Similar steps have been applied to other WFDs.

The weight boxes fusion ensemble structure is an ensemble model initially developed by Solovyev et al., which uses the confidence scores of all bounding boxes proposed to develop the averaged boxes [35]. The basic structure of the weight boxes fusion ensemble method is shown in Figure 6 below.

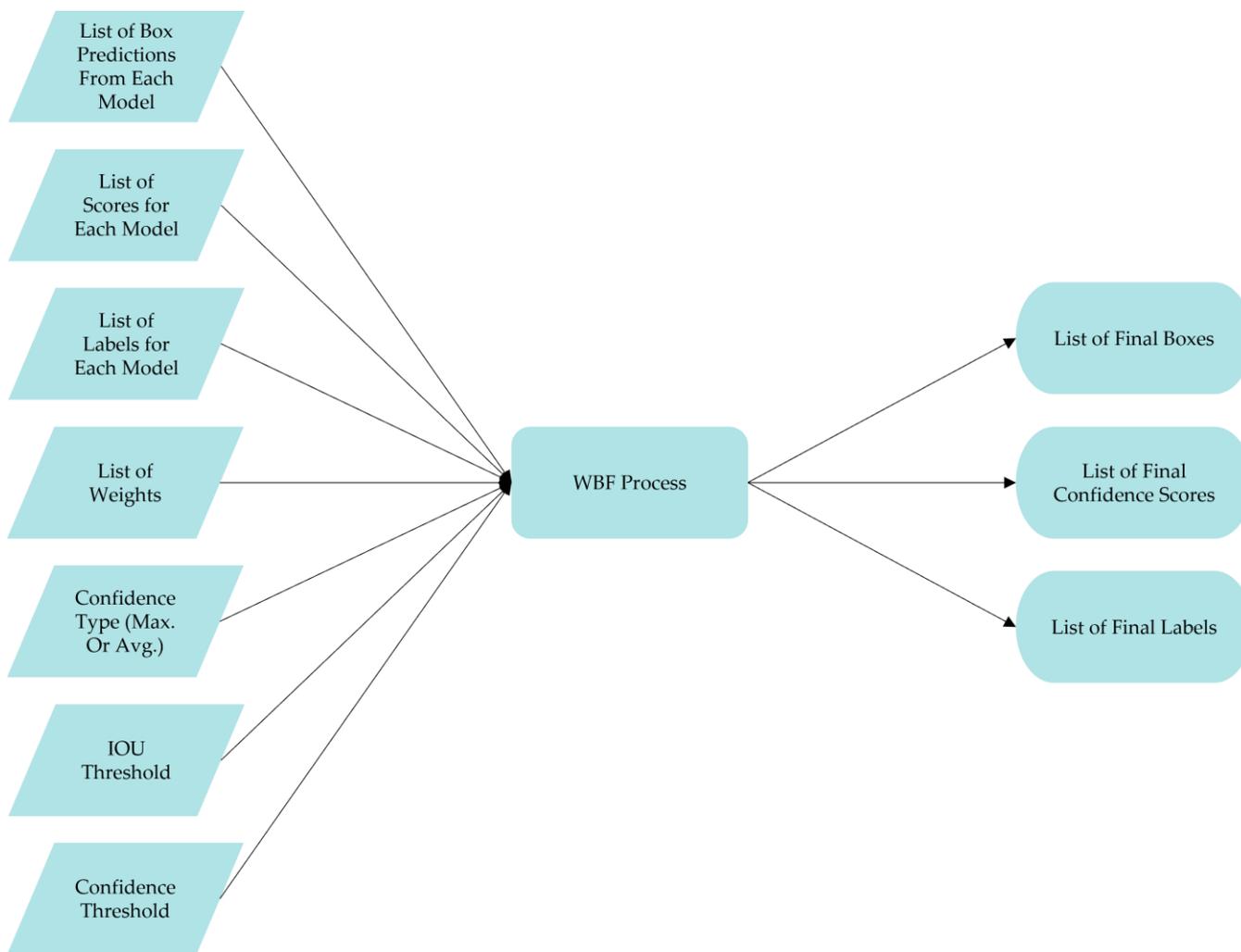

**Figure 6.** Weight boxes fusion ensemble.

The scheme of the main WFD ensemble models used within the scope of the study is displayed in Figure 7 below.

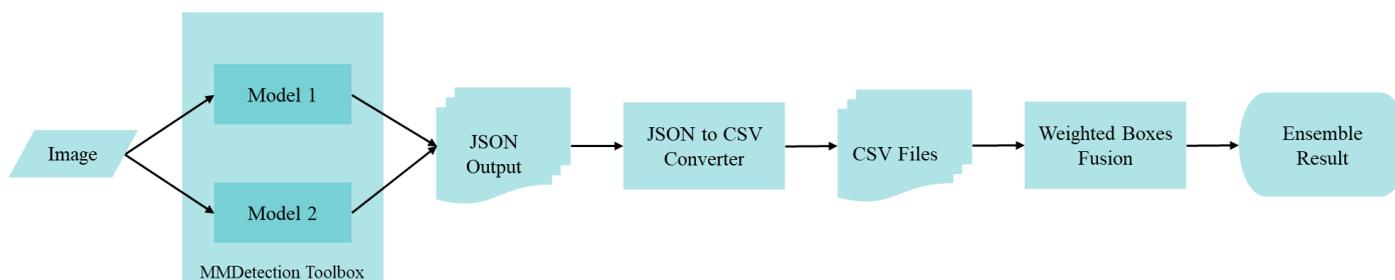

**Figure 7.** WFD ensemble models based on weight boxes fusion.

With the WFD ensemble model based on weight boxes fusion in Figure 7 above, firstly, JSON outputs for fracture detection are obtained, using the MMDetection Toolbox. Subsequently, JSON files are converted into CSV format. In these files, ensemble results are achieved by performing weight boxes fusion with different weight coefficients varying from 1 to n, taking into account the number of models (n) used for detection.

The scheme of the wrist fracture detection-combo (WFD-C) ensemble model developed in this study for fracture detection in wrist X-ray images is presented in Figure 8 below. In the WFD-C model, another weight boxes fusion-based ensemble procedure was performed using the previously achieved WFD-1, 2, 3, 4 and 5 models. The submodels were used in 6 different ensemble models in total, and the corresponding weight coefficients thereof are displayed in Table 2 below.

**Table 2.** Submodels and weight coefficients of WFD ensemble models.

| Ensemble Models | Model-1 | Model-2 | Model-3 | Model-4 | Model-5 | Weight Coefficients |
|---|---|---|---|---|---|---|
| WFD-1 (single stage) | RegNet | FSAF | RetinaNet | SABL Rt.Net | PAA | (1, 5, 5, 5, 5) |
| WFD-2 (two stage) | DCN | SABL Fs. R-CNN | - | - | - | (2, 2) |
| WFD-3 (AP50) | RegNet | PAA | FSAF | Libra Rt.Net | - | (3, 3, 3, 4) |
| WFD-4 (AR) | RegNet | PAA | FSAF | SABL Rt.Net | - | (1, 3, 3, 1) |
| WFD-5 (LRP-opt.) | FSAF | Fs. R-CNN | DCN | Libra Rt.Net | RetinaNet | (2, 1, 4, 2, 3) |
| WFD-C (combo) | WFD-1 | WFD-3 | WFD-4 | WFD-5 | WFD-2 | (4, 4, 3, 5, 5) |

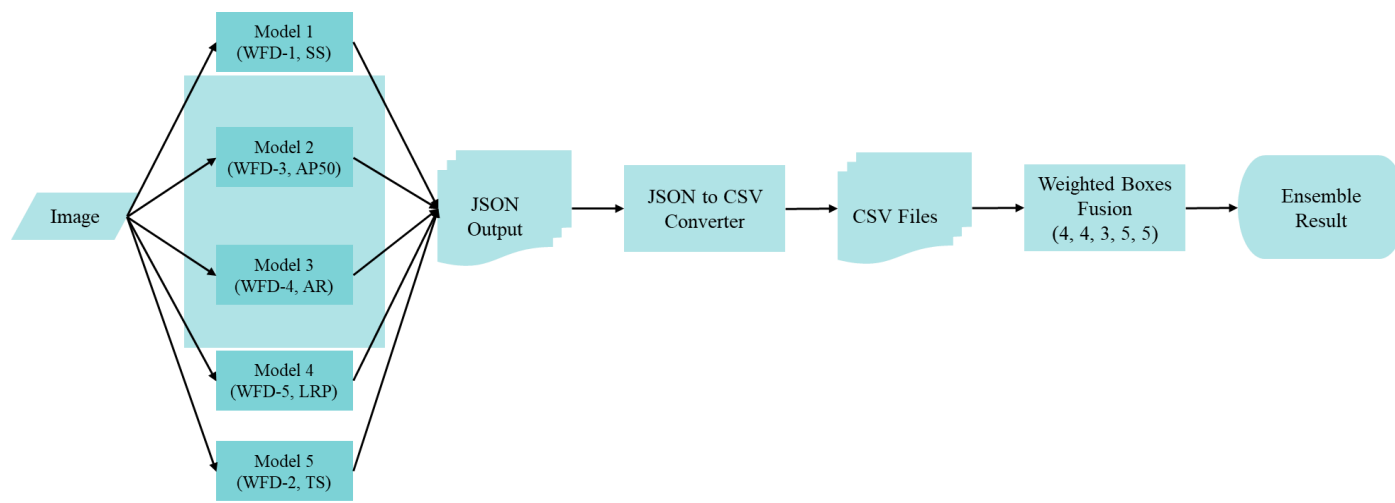

**Figure 8.** Proposed WFD-C ensemble models based on weight boxes fusion.

## 4. Experiments

Deep-learning-based object detection models were used in this study, in which bone fracture detection was performed on wrist X-ray images. Using a dataset collected from Gazi University Hospital, the first procedures performed were the labeling of bone fractures, data preprocessing and data augmentation. Subsequently, 20 different detection procedures were performed using a number of different object detection models with various backbones, which are based on deep-learning. Taking into account the results achieved herein, ensemble models were developed to further improve the detection results. The bone fracture detection models used and proposed for wrist X-ray images are presented in Figure 9 below.

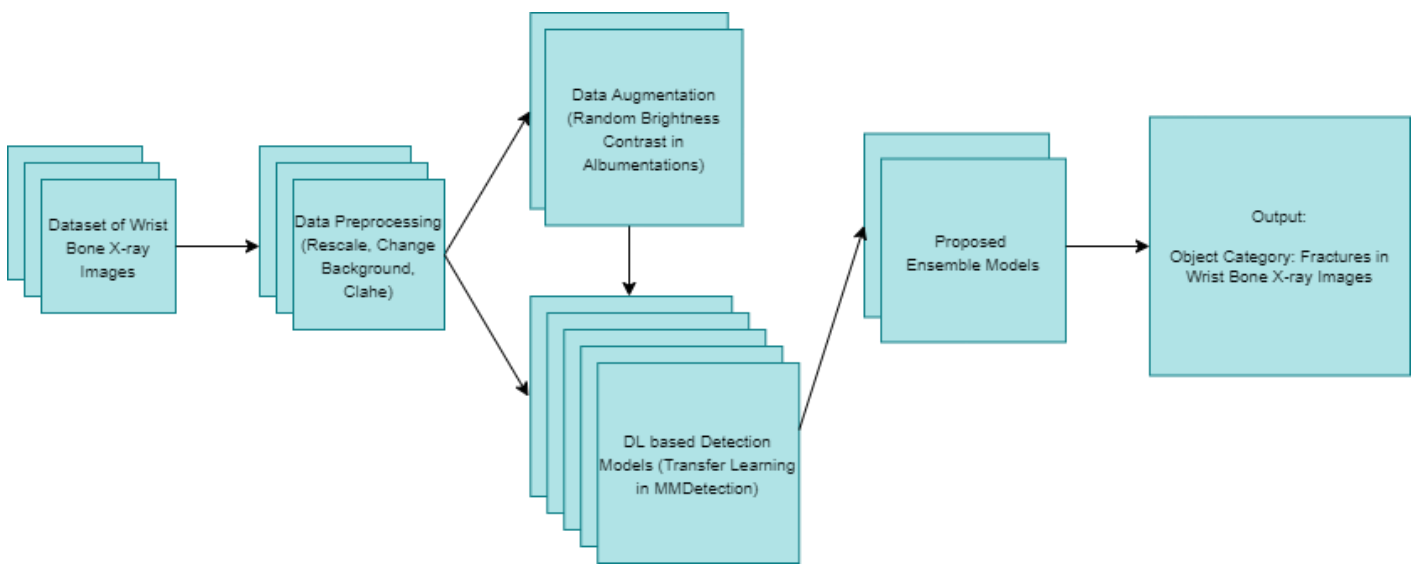

**Figure 9.** The proposed models for fracture detection of wrist X-ray images.

*4.1. Dataset and Labeling of Wrist X-ray Images*

The wrist X-ray images collected from Gazi University Hospital were used within the scope of the study. The official document (date and number: 27.04.2020-E.51150) confirming that collection and use of this dataset does not constitute any ethical inconvenience was provided by the Measurement and Evaluation Ethics Working Sub-Group of Gazi University. The physicians Dr. Murat Çiçeklidağ, Assoc. Prof. Dr. Tolga Tolunay and Prof. Dr. Nil Tokgöz, who work at Gazi University Hospital, provided assistance in data collection and labeling of the fracture area. The number of wrist X-ray images collected from the Radiology Department of the hospital is 542, and the image format is Digital Imaging and Communications in Medicine (DICOM). This dataset belongs to patients between the years 2010–2020. All of these patients are patients who presented to the emergency department. All images in the dataset were obtained from the same X-ray machine. The X-ray machine model used is the Samsung GC70. There is a heterogeneous distribution of both right-wrist and left-wrist images in the dataset. No distinction was made in the treatment and use process of these. According to the results obtained using the Python pydicom library, there was a total of 275 patients. Of these, 134 females and 141 males were available. The average age was 44.99. There were a total of 21 individuals aged 12 and under. A total of 92.37% of all patients were adults and 7.63% were under 12 years old. In order to use the collected data in CNN-based object detection models, the format was converted from DICOM format to 3-channel png format. Pydicom library was used to read and extract information from images taken in DICOM format. Images in DICOM format were converted to grayscale PNG format using the library. After certain operations were made on the images in PNG format, normalization was made within the framework in the object detection network, and training and testing were carried out by converting them to RGB format in order to perform operations such as coloring on the image in the future studies. Moreover, the graphical image annotation tool titled LabelImg [36] was used to label the areas of fracture in the images.

An X-ray device only was used for imaging the wrists of the patients. The X-ray images obtained from this device were also examined by three physicians who are experts in their fields. For this reason, there was no need for a different imaging technique such as CT or MRI. The labeling of the fractures was jointly examined by one radiologist and two orthopedists.

Each of the images in the dataset was examined by one radiologist (Nil Tokgöz) working in the Radiology Department at Gazi University Hospital and two orthopedist (Murat Çiçekdağ, Tolga Tolunay) working in the Orthopedics and Traumatology

Department (Murat Çiçekdağ, Tolga Tolunay) and the fractures were detected. Subsequently, the use of the labeling library was explained to the physicians in detail and the labeling was provided by the physicians. The fracture areas examined and labeled by the physicians in the bone X-ray images used within the scope of the study belong only to the radius and ulna bones.

All of the wrist images taken from Gazi University Hospital consist of fracture (abnormal, unhealthy, positive) images. Therefore, there is at least one fracture in each of the images in the datasets. There are a total of 569 fractures in 542 wrist images used within the scope of the study. The distribution of the 569 fractures in the dataset is as follows: there are 459 labels in 434 training data, 55 labels in 54 validation data and 56 labels in 54 test data.

Figure 10 below shows the distribution of the wrist X-ray dataset, which initially had different resolutions in terms of quantity and percentage, as a training, validation and test dataset. The images in the training, validation and test dataset belong to different patients. Therefore, if the images of a patient are more than one, they are included in the same group dataset. The number of patients with both hands fractures was 10 in the train dataset, 1 in the test dataset, and none in the validation dataset. In addition, out of 434 images in the training dataset, 28 were pairs (right and left hands), 187 belong to the right hand, and 219 belong to the left hand. In the validation dataset, there are 3 pairs and 24 right and 27 left hand images out of 54 images In the test dataset, there are 7 pairs and 22 right and 25 left hand images out of 54 images. Of the total images, 7% are pair, 43% are right-handed and 50% are left-handed images. All of the wrist X-ray images used in the study and taken from Gazi University Hospital are abnormal (positive, unhealthy, fracture) images. No normal (negative) images were obtained from the hospital. The distribution of our wrist X-ray abnormal (fracture) image dataset, consisting of 542 images, is 80% training, 10% validation, 10% test. All details of the wrist fracture dataset are given in Table 3.

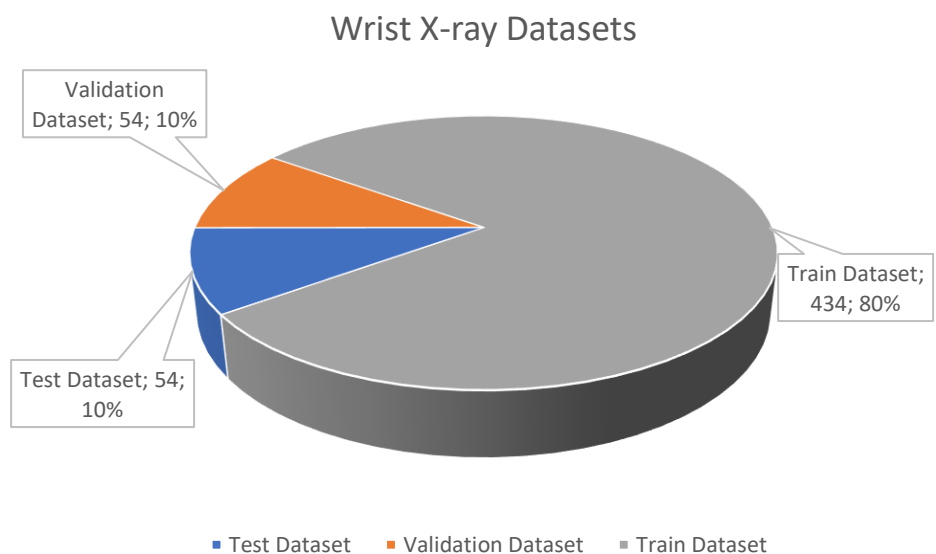

**Figure 10.** The wrist X-ray dataset.

**Table 3.** Details of wrist fracture dataset.

| X-ray Device | Samsung GC70 | Time Period of Collection | 2010–2020 |
|---|---|---|---|
| X-ray images and total fractures | 542, 569 | Specialist physicians | 1 radiologist, 2 orthopedists |
| Dataset (train, validation, test) | %80, %10, %10 | Fracture types | radius and ulna |
| Training data and fractures | 434, 459 | Patients (female, male, total) | 134, 141, 275 |
| Validation data and fractures | 54, 55 | Average age, pediatrics, adult | 45, 21, 254 |
| Test data and fractures | 54, 56 | Patients with fractures in both wrists (train, validation, test, total) | 10, 0, 1, 11 |

*4.2. Data Preprocessing of Wrist X-ray Images*

When the wrist X-ray image dataset used within the scope of the study and collected from Gazi University Hospital was examined after the format was converted from DICOM to 3-channel png format, a difference in their background was observed.

When the wrist X-ray images used within the scope of the study were examined, especially after they were converted from DICOM format to png format, it was understood that there was no standard in image sizes, but that they were different sizes. To the extent that the object detection models used and local PC hardware supported it, the wrist images in the images were first manually cropped to include the hand and wrist parts. Afterward, all input image sizes were rescaled to 800 × 800 × 3, so as not to affect the fracture detection process to the extent supported by local PC hardware.

The dominant color displayed by the background of the image was identified by using the binary K-means algorithm to recover the difference in the backgrounds of the images. Following this procedure, the images were inverted (white if black, black if white) if the dominant color is white, and the background was converted into black and the foreground into white. This procedure ensured that all images used in the dataset were in a certain color format. After adjustment of the color format, contrast-limited adaptive histogram equalization (CLAHE) with 11 × 11 grid size and 7.0 clip limit parameters was applied on the images. Figure 11 below shows the sample images of the wrist X-ray image dataset achieved as the result of the data preprocessing steps.

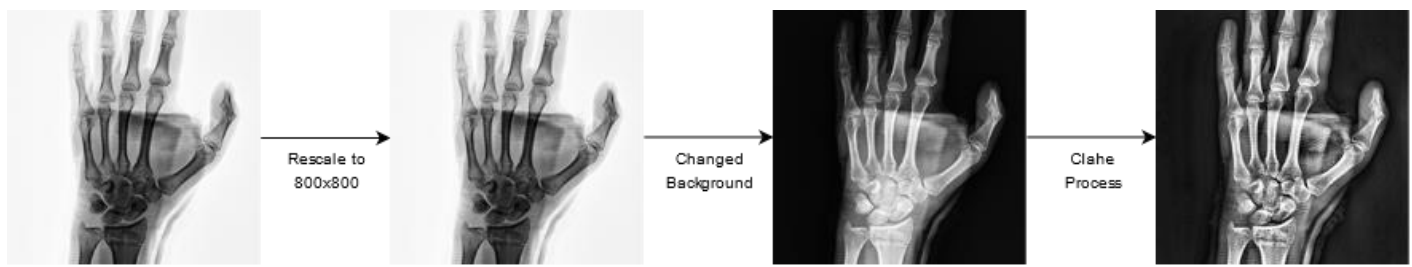

**Figure 11.** The wrist X-ray images after data preprocessing.

*4.3. Data Augmentation of Wrist X-ray Images*

The amount of data is essential for network training with deep-learning-based object detection models. Therefore, data augmentation was performed within the scope of this study in order to make the best of network training and to achieve high scores for fracture detection. As for the initial dataset, various augmentations were tried using the Albumentations [37] library. Albumentations is a flexible and fast image augmentation python library that can be used in different computer vision tasks that include object detection, classification and segmentation, which are particularly deep-learning-based open source projects [37]. In augmentations performed using this library, random brightness contrast, sharpness, noise, gamma, gaussian blur and median blur were used. Experiments of augmentation were conducted using these six different methods used in these procedures, either individually or together. Upon analysis of the detection procedures performed with augmentation, methods that have a negative impact on the result of detection, methods with no impact or the methods inclined to augmentation were identified. Based on the experiments, it was concluded that random brightness contrast augmentation made the greatest contribution for detection of fractures in the Albumentations library. This type of augmentation was used in addition to the random flip ratio 0.5 augmentation available in the training stage of the models in the MMdetection tool used for detection. Samples of the images obtained as the result of the augmentation steps are presented in Figure 12 below.

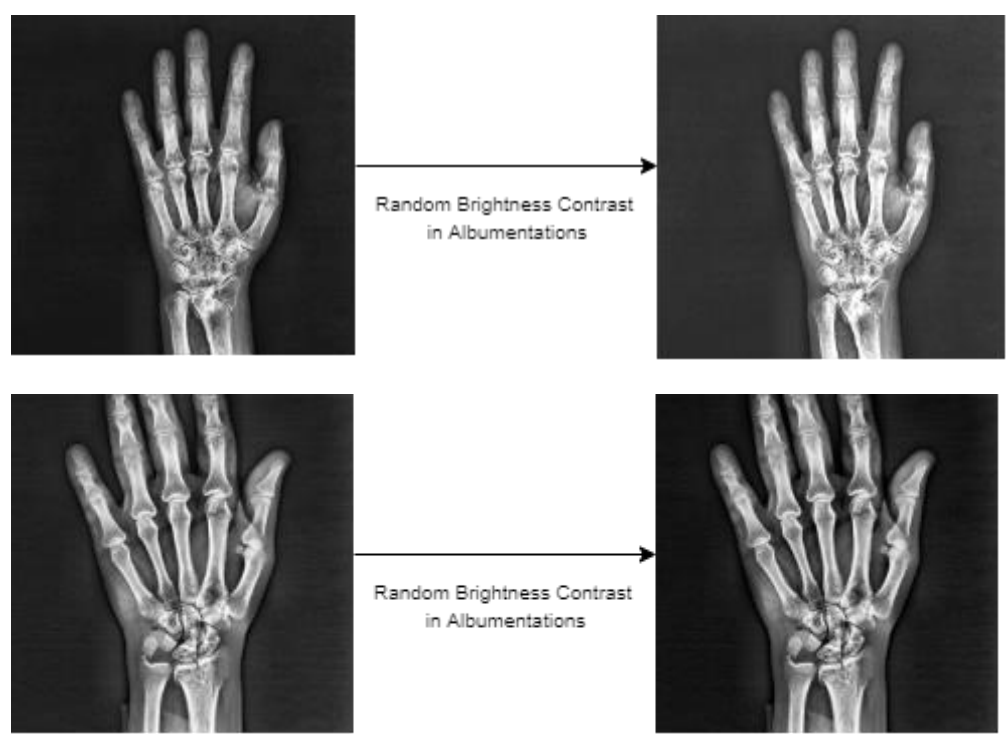

**Figure 12.** The wrist X-ray images after data augmentation.

*4.4. Fracture Detection Results*

Offline local PCs and Google Colab were used as hardware in fracture detection procedures performed using deep-learning-based object detection models in wrist X-ray images. The graphics cards of the hardware used for fracture detection performed in these systems are as follows: 4 GB nVIDIA GTX1650 and GTX1050 for the offline local PCs, and nVIDIA Tesla T4 16 GB GDDR6 for Google Colab. In addition to these, Albumentations, LRP error [38], review object detection metrics [39], weight boxes fusion [35] packages were used. MMDetection [40] toolbox was used for object detection procedures. MMDetection is an open source object detection toolbox based on PyTorch containing many different deep-learning-based object detection models, including single-stage, two-stage and multi-stage, which is used in object detection and instance segmentation procedures in particular [40]. The program codes written for this study using PyTorch-based deep learning libraries are publicly available at https://github.com/fatihuysal88/wrist-d (accessed on 30 December 2021). The following were used in all fracture detection procedures performed using deep-learning-based object detection models in Figure 2: learning rate 0.001, epoch number 40, optimizer SGD, momentum 0.9, Cross Entropy Loss or Focal Loss as class loss function and Smooth L1

Loss, IoU Loss, Balanced L1 Loss or L1 Loss as bounding box loss function. In order to improve the success of network learning used in all these models, the learning rate value was not kept constant. During the 40 epochs of training, the learning rate decreased 10 times in the 5th, 10th, 15th, 25th and 35th epochs, respectively. The parameters used in the ensemble models developed specifically for the study are skip box threshold 0.3, intersection over union threshold 0.5 and limit boxes 6000. Although not defaulted in the training parameters, it has been chosen on a per-study basis to create an overall structure similar to the suggested configurations. Since the number of epochs depends on the complexity of the model, it has been kept high and has been chosen as 40 in order to appeal to each model. Accordingly, choosing a higher epoch number has allowed the selection of a lower and generally accepted learning rate. In addition, augmentation steps and other optimization settings are the same in all models.

### 4.4.1. Evaluation Metrics

In order to evaluate the results achieved in object detection problems in the most efficient way, average precision (AP), average recall (AR), precision-recall curve, ground truth-prediction comparison on image and localization recall precision (LRP) error [38] scores must be obtained. Precision and Recall scores depend on the threshold value of Intersection Over Union (IOU), as well as the True Positive (TP), False Positive (FP) and False Negative (FN) values. IOU suggests the ratio (part) of the overlapping area to union area of the predicted bounding box and the ground-truth bounding box. In order for classification of the detection result to be true or false, a threshold value must be specified. If the IOU score is greater than or equal to the threshold, the detection result is considered to be true, in other cases (if the IOU is less than the threshold), the detection result is considered to be false. TP refers to the correct detection of ground-truth bounding. FP is the misplaced detection of an existing object or the false detection of a non-existing object. FN stands for failure of detection of ground-truth bounding. Although it is used in classification, there is no value True Negative (TN) in object detection problems. This is because the TN value cannot be used as there are an infinite number of bounding boxes that should not be detected in any image. Precision (P) refers to the ratio of TP to all detections (TP + FP). Recall (R) is the ratio of TP to all ground-truths (TP + FN). AP is the average precision over all unique recall levels and refers to the area under the precision-recall curve. AR is the average recall over the whole IOU in the [0.5, 1.0] interval and is calculated as twice the area under the recall-IOU curve. LRP error is a metric proposed by Oksuz et al., in 2018 that is applicable to all object detection tasks [41]. This error metric is also an alternative to AP for key point detection, instance segmentation and object detection [38]. Table 4 below explains the meanings of IOU, TP, TN, FP and FN values and how the P, R, AP, AR and optimal LRP are calculated.

**Table 4.** Evaluation metrics definition and calculations.

| Evaluation Metrics | Definition and Calculations |
|---|---|
| Intersection over Union (IOU) | area ($BBox_p \cap Bbox_g$)/area ($BBox_p \cup Bbox_g$) |
| True Positive (TP) | $IUO \geq 0.5$ |
| False Positive (FP) | $IUO < 0.5$ |
| False Negative (FN) | failing to detect $Bbox_g$ |
| True Negative (TN) | can't be used (infinite) |
| Precision (P) | TP/(TP + FP) = TP/all detections |
| Recall (R) | TP/(TP + FN) = TP/all ground truths |
| Average Precision (AP) | under area of P-R curve |
| Average Recall (AR) | twice the under area of R-IOU curve |
| Optimal Localization Recall Precision (oLRP) | minimum achievable average matching error over the confidence scores |

### 4.4.2. Fracture Detection Results of 20 DL-Based Models

For fracture detection in wrist X-ray images, a total of 20 fracture detection procedures were performed, with and without augmentation, in 10 different deep-learning-based models. Figures 13 and 14 and Tables 5–7 present the following results for fracture detection procedures performed with each model: train bbox loss (TB_Loss), train loss (T_Loss) and training time (TT) values for the training phase, the epoch where the highest validation accuracy is achieved for AP50, AR, oLRP, $oLRP_{Loc}$, $oLRP_{FP}$ and $oLRP_{FN}$ values for the validation phase.

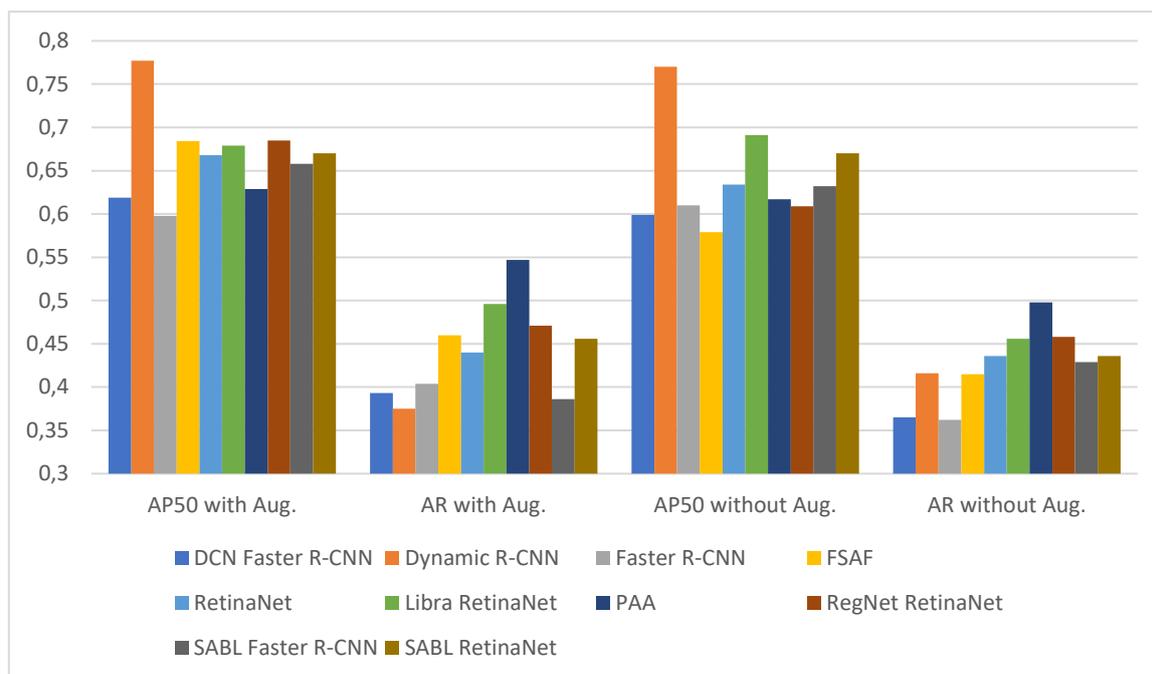

**Figure 13.** Validation AP50 and AR results of detection models.

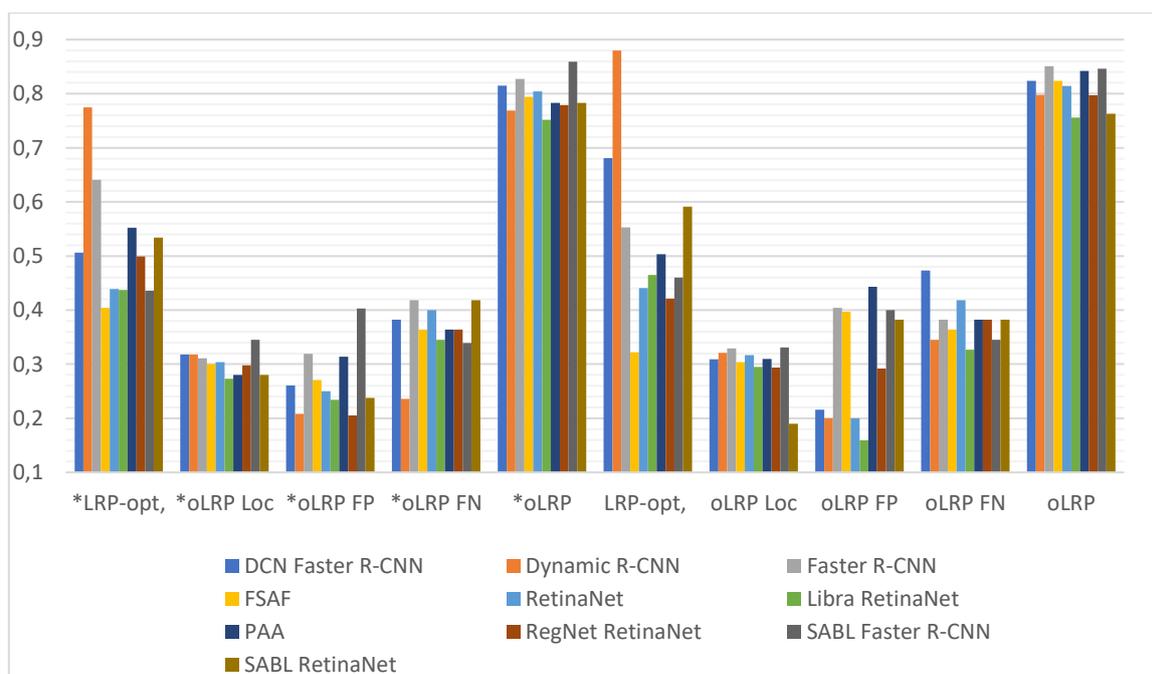

**Figure 14.** Validation LRP-optimal threshold ($LRP_t$), $oLRP_{Loc}$ ($oLRP_L$), $oLRP_{FP}$ and $oLRP_{FN}$ results of detection models. (* with Aug.).

**Table 5.** Training loss and epoch results for the highest validation AP50 in detection models.

| Models | Without Augmentation | | | | With Augmentation | | | |
|---|---|---|---|---|---|---|---|---|
| | TB_Loss | T_Loss | TT | Best Epoch | TB_Loss | T_Loss | TT | Best Epoch |
| DCN Faster R-CNN | 0.0938 | 0.1821 | 174 | 6 | 0.0969 | 0.1839 | 174 | 5 |
| Dynamic R-CNN | 0.2978 | 0.5033 | 139 | 8 | 0.2664 | 0.4605 | 156 | 10 |
| Faster R-CNN | 0.0873 | 0.1573 | 126 | 10 | 0.0864 | 0.1634 | 137 | 12 |
| FSAF | 0.3142 | 0.5419 | 130 | 7 | 0.2605 | 0.4718 | 128 | 8 |
| RetinaNet | 0.3103 | 0.4987 | 120 | 16 | 0.348 | 0.576 | 120 | 8 |
| Libra RetinaNet | 0.5033 | 0.7278 | 131 | 8 | 0.5374 | 0.7813 | 130 | 6 |
| PAA | 0.3256 | 0.8387 | 127 | 6 | 0.33 | 0.834 | 104 | 7 |
| RegNet RetinaNet | 0.315 | 0.543 | 256 | 8 | 0.313 | 0.539 | 225 | 7 |
| SABL Faster R-CNN | 0.0536 | 0.2244 | 243 | 8 | 0.0501 | 0.2140 | 212 | 12 |
| SABL RetinaNet | 0.1034 | 0.3817 | 132 | 21 | 0.1857 | 0.6559 | 123 | 6 |

**Table 6.** Validation AP50 and AR results of detection models.

| Models | Without Augmentation | | With Augmentation | |
|---|---|---|---|---|
| | AP50 | AR | AP50 | AR |
| DCN Faster R-CNN | 0.599 | 0.365 | 0.619 | 0.393 |
| Dynamic R-CNN | 0.77 | 0.416 | 0.777 | 0.375 |
| Faster R-CNN | 0.61 | 0.362 | 0.598 | 0.404 |
| FSAF | 0.579 | 0.415 | 0.684 | 0.46 |
| RetinaNet | 0.634 | 0.436 | 0.668 | 0.44 |
| Libra RetinaNet | 0.691 | 0.456 | 0.679 | 0.496 |
| PAA | 0.617 | 0.498 | 0.629 | 0.547 |
| RegNet RetinaNet | 0.609 | 0.458 | 0.685 | 0.471 |
| SABL Faster R-CNN | 0.632 | 0.429 | 0.658 | 0.386 |
| SABL RetinaNet | 0.67 | 0.436 | 0.67 | 0.456 |

**Table 7.** Validation LRP-optimal threshold ($LRP_t$), $oLRP_{Loc}$ ($oLRP_L$), $oLRP_{FP}$ and $oLRP_{FN}$ results of detection models.

| Models | Without Augmentation | | | | | With Augmentation | | | | |
|---|---|---|---|---|---|---|---|---|---|---|
| | $LRP_t$ | $oLRP_L$ | $oLRP_{FP}$ | $oLRP_{FN}$ | oLRP | $LRP_t$ | $oLRP_L$ | $oLRP_{FP}$ | $oLRP_{FN}$ | oLRP |
| DCN Faster R-CNN | 0.681 | 0.309 | 0.216 | 0.473 | 0.824 | 0.506 | 0.318 | 0.261 | 0.382 | 0.815 |
| Dynamic R-CNN | 0.88 | 0.321 | 0.2 | 0.345 | 0.798 | 0.775 | 0.318 | 0.208 | 0.236 | 0.769 |
| Faster R-CNN | 0.553 | 0.329 | 0.404 | 0.382 | 0.851 | 0.641 | 0.311 | 0.319 | 0.418 | 0.827 |
| FSAF | 0.322 | 0.304 | 0.397 | 0.364 | 0.824 | 0.404 | 0.3 | 0.271 | 0.364 | 0.794 |
| RetinaNet | 0.441 | 0.317 | 0.2 | 0.418 | 0.814 | 0.439 | 0.304 | 0.25 | 0.4 | 0.804 |
| Libra RetinaNet | 0.465 | 0.295 | 0.159 | 0.327 | 0.756 | 0.437 | 0.273 | 0.234 | 0.345 | 0.752 |
| PAA | 0.503 | 0.31 | 0.443 | 0.382 | 0.842 | 0.552 | 0.28 | 0.314 | 0.364 | 0.783 |
| RegNet RetinaNet | 0.421 | 0.294 | 0.292 | 0.382 | 0.797 | 0.499 | 0.298 | 0.205 | 0.364 | 0.779 |
| SABL Faster R-CNN | 0.46 | 0.331 | 0.4 | 0.345 | 0.846 | 0.436 | 0.345 | 0.403 | 0.339 | 0.859 |
| SABL RetinaNet | 0.591 | 0.19 | 0.382 | 0.382 | 0.763 | 0.534 | 0.28 | 0.238 | 0.418 | 0.783 |

It can be observed upon examination of Table 5 that the number of epochs with the highest validation accuracy varies between 6–21 in models without augmentation and 5–12 in models with augmentation, and that the lowest loss values achieved from different training times in models both with and without augmentation is achieved in SABL Faster R-CNN for train bbox loss and in DCN Faster R-CNN model for train loss. The analysis of the values in Table 6 and the graph in Figure 13 suggest that the best AP50 scores in validation were obtained in Dynamic R-CNN models with/without augmentation among the models used for detection. The Figure 14 and Table 7 show that the lowest LRP optimal threshold values for validation are available in FSAF in all models.

The results of the tests performed with the test data following the training with and without augmentation performed with each model for fracture detection are presented in detail in Figures 15 and 16 and Tables 8 and 9 below.

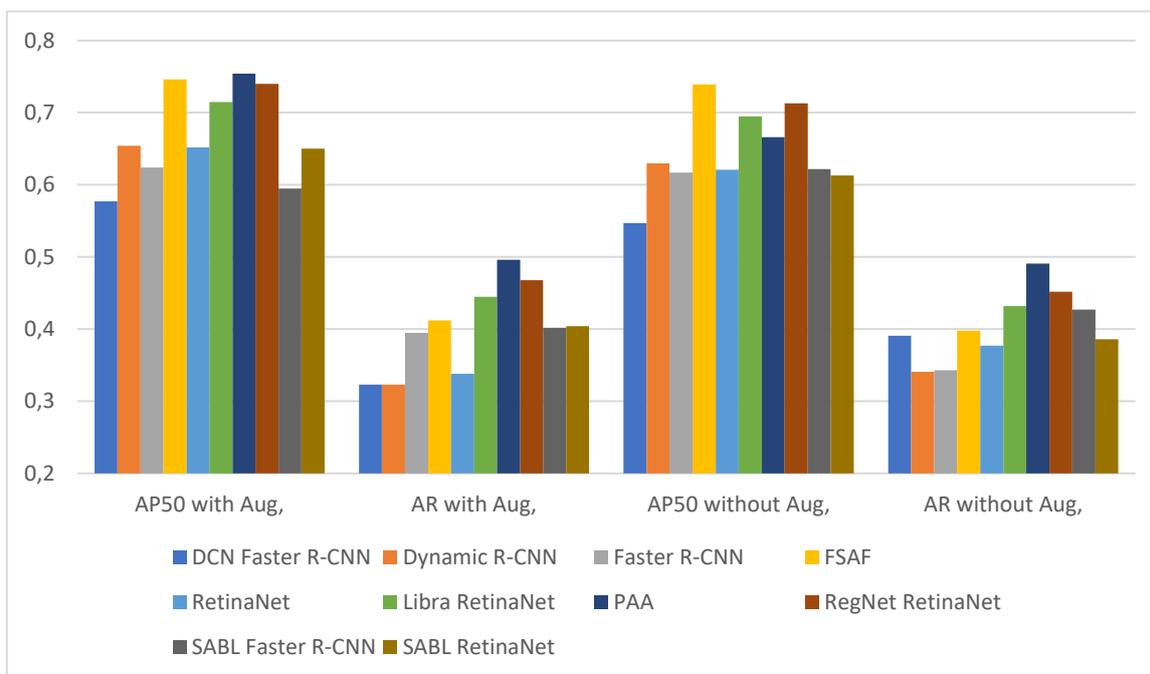

**Figure 15.** Test AP50 and AR results of detection models.

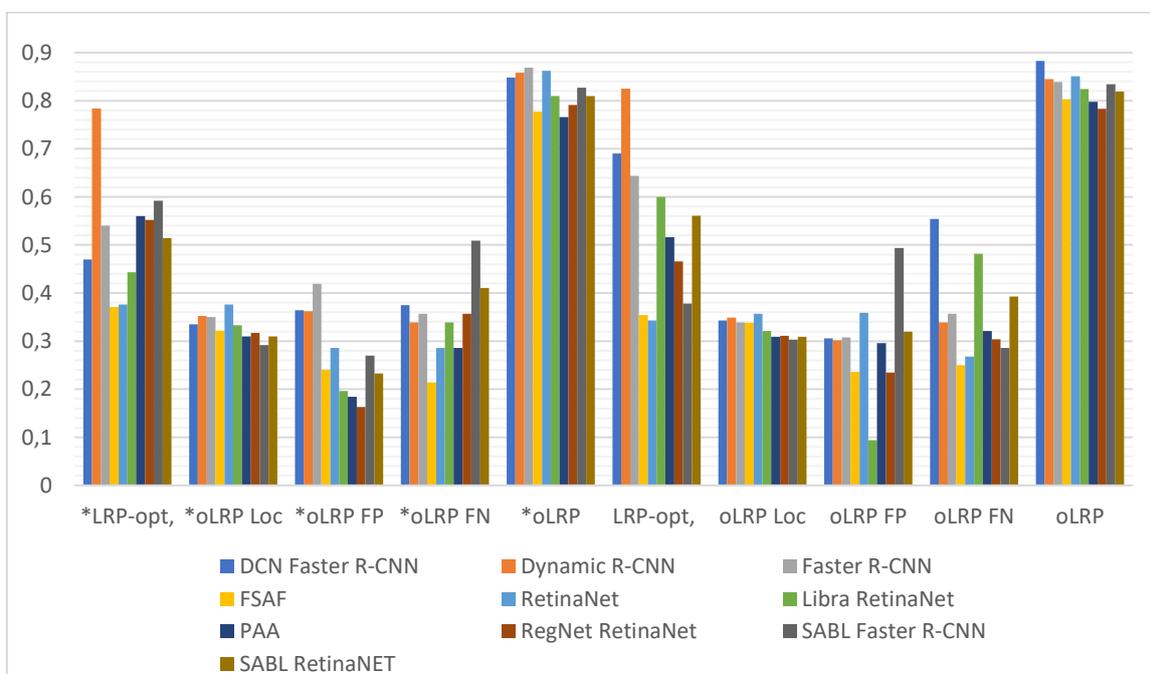

**Figure 16.** Test LRP-optimal threshold (LRP$_t$), oLRP$_{Loc}$ (oLRP$_L$), oLRP$_{FP}$ and oLRP$_{FN}$ results of detection models. (* with Aug.).

**Table 8.** Test AP50 and AR results of detection models.

| Models | Without Augmentation | | With Augmentation | |
|---|---|---|---|---|
| | AP50 | AR | AP50 | AR |
| DCN Faster R-CNN | 0.547 | 0.391 | 0.577 | 0.323 |
| Dynamic R-CNN | 0.63 | 0.341 | 0.654 | 0.323 |
| Faster R-CNN | 0.617 | 0.343 | 0.624 | 0.395 |
| FSAF | 0.739 | 0.398 | 0.746 | 0.412 |
| RetinaNet | 0.621 | 0.377 | 0.652 | 0.338 |
| Libra RetinaNet | 0.695 | 0.432 | 0.715 | 0.445 |
| PAA | 0.666 | 0.491 | 0.754 | 0.496 |
| RegNet RetinaNet | 0.713 | 0.452 | 0.74 | 0.468 |
| SABL Faster R-CNN | 0.622 | 0.427 | 0.595 | 0.402 |
| SABL RetinaNet | 0.613 | 0.386 | 0.65 | 0.404 |

**Table 9.** Test LRP-optimal threshold ($LRP_t$), $oLRP_{Loc}$ ($oLRP_L$), $oLRP_{FP}$ and $oLRP_{FN}$ results of detection models.

| Models | Without Augmentation | | | | | With Augmentation | | | | |
|---|---|---|---|---|---|---|---|---|---|---|
| | $LRP_t$ | $oLRP_L$ | $oLRP_{FP}$ | $oLRP_{FN}$ | oLRP | $LRP_t$ | $oLRP_L$ | $oLRP_{FP}$ | $oLRP_{FN}$ | oLRP |
| DCN Faster R-CNN | 0.69 | 0.343 | 0.306 | 0.554 | 0.883 | 0.47 | 0.335 | 0.364 | 0.375 | 0.848 |
| Dynamic R-CNN | 0.825 | 0.349 | 0.302 | 0.339 | 0.845 | 0.784 | 0.352 | 0.362 | 0.339 | 0.858 |
| Faster R-CNN | 0.644 | 0.339 | 0.308 | 0.357 | 0.839 | 0.54 | 0.35 | 0.419 | 0.357 | 0.869 |
| FSAF | 0.354 | 0.338 | 0.236 | 0.25 | 0.803 | 0.371 | 0.322 | 0.241 | 0.214 | 0.777 |
| RetinaNet | 0.343 | 0.357 | 0.359 | 0.268 | 0.851 | 0.376 | 0.376 | 0.286 | 0.286 | 0.862 |
| Libra RetinaNet | 0.6 | 0.321 | 0.094 | 0.482 | 0.824 | 0.443 | 0.333 | 0.196 | 0.339 | 0.81 |
| PAA | 0.516 | 0.309 | 0.296 | 0.321 | 0.798 | 0.56 | 0.310 | 0.184 | 0.286 | 0.766 |
| RegNet RetinaNet | 0.466 | 0.311 | 0.235 | 0.304 | 0.783 | 0.552 | 0.317 | 0.163 | 0.357 | 0.791 |
| SABL Faster R-CNN | 0.378 | 0.303 | 0.494 | 0.286 | 0.834 | 0.592 | 0.292 | 0.27 | 0.509 | 0.827 |
| SABL RetinaNet | 0.561 | 0.309 | 0.32 | 0.393 | 0.819 | 0.514 | 0.31 | 0.233 | 0.411 | 0.81 |

It is observed upon examination of Table 8 and Figure 15 that among all the models used for fracture detection in wrist X-ray images, the highest AP50 score obtained on the test data was 0.754 in PAA model with augmentation. Figure 16 and Table 9 indicate that the lowest values of LRP optimal threshold, oLRPLoc, oLRPFP, oLRPFN and oLRP in models with augmentation were obtained in the FSAF, SABL Faster R-CNN, RegNet RetinaNet, PAA/RetinaNet and PAA models, respectively.

The bounding box outputs achieved from the fracture detection performed with the PAA model with the best AP50 score are displayed in Figure 17 for right/left hand as a sample in the dataset, and the precision-recall graph is provided in Figure 18.

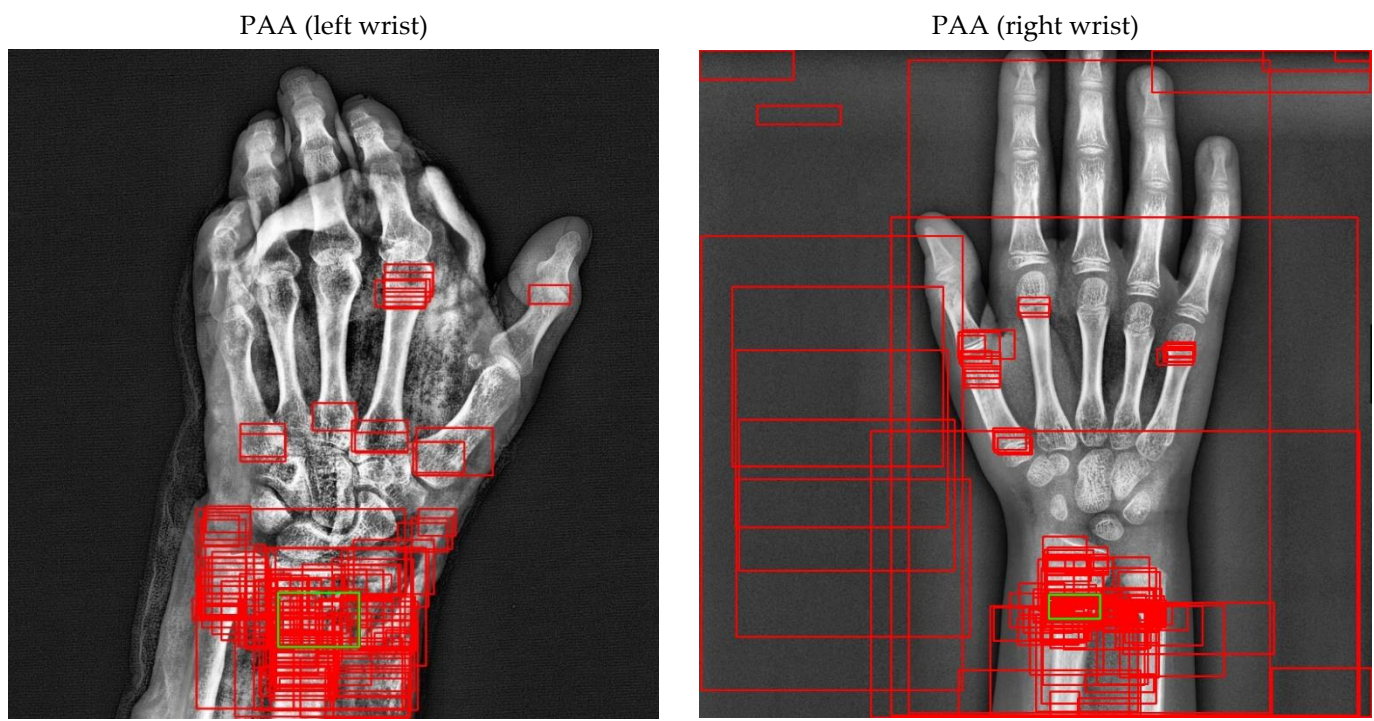

**Figure 17.** Sample of left/right wrist fracture results [ground-truth bounding box (green), predicted bounding box (red)] for PAA (Best score of 20 models).

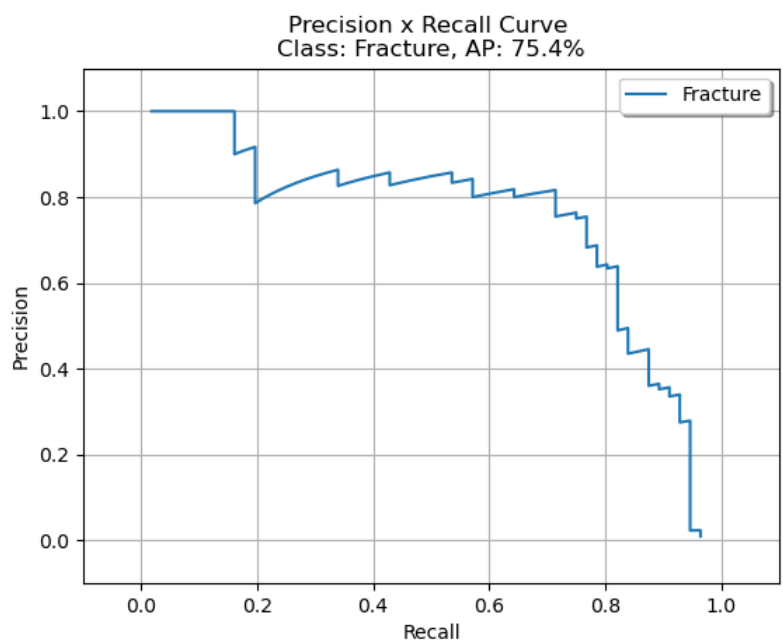

**Figure 18.** Precision-recall curve of PAA (best score of 20 models).

In addition to the 20 procedures of fracture detection performed on wrist X-ray images based on deep learning provided in this section and thereof, the outputs of the ensemble models proposed to improve the detection results further are explained in the next section.

### 4.4.3. Fracture Detection Results of Proposed Models

Based on the results of 20 models based on deep learning in which fracture detection was performed in wrist X-ray images as explained in the previous section, ensemble models were developed, thus leading to an improvement in the detection results. AP50, AR and LRP scores achieved from six different WFD-based ensemble models are presented in Figure 19 and Table 10 below.

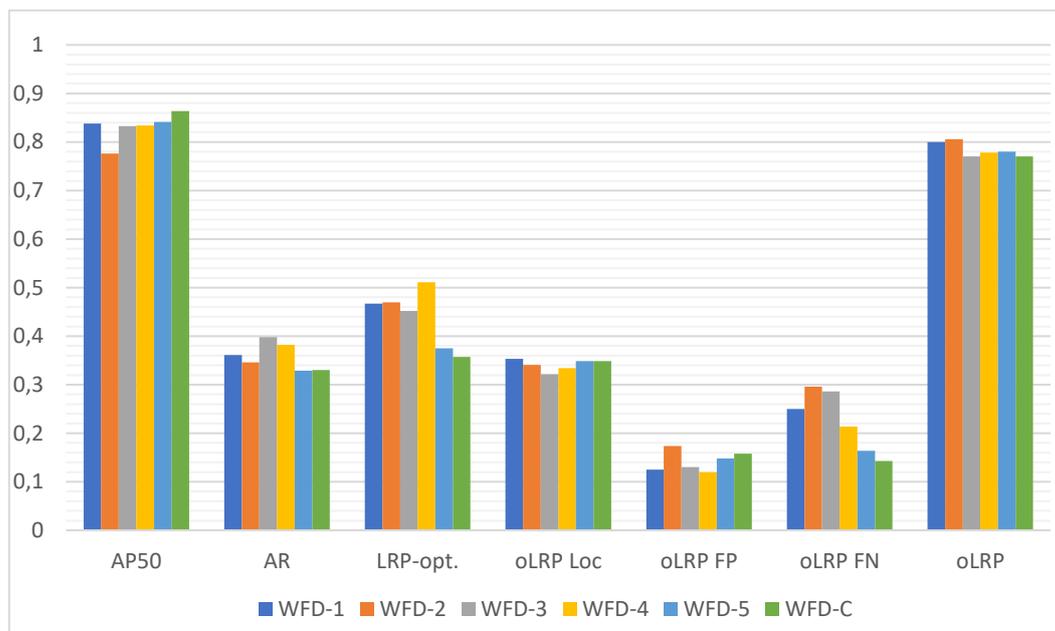

**Figure 19.** Results (AP50, AR, LRP-optimal threshold (LRPt), oLRPLoc (oLRPL), oLRPFP and oLRPFN) of Ensemble Model (WFD-1, WFD-2, WFD-3, WFD-4, WFD-5, WFD-C).

**Table 10.** Results (AP50, AR, LRP-optimal threshold (LRP$_t$), oLRP$_{Loc}$ (oLRP$_L$), oLRP$_{FP}$ and oLRP$_{FN}$) of Ensemble Model (WFD-1, WFD-2, WFD-3, WFD-4, WFD-5, WFD-C).

| Ensemble Models | AP50 | AR | LRP$_t$ | oLRP$_L$ | oLRP$_{FP}$ | oLRP$_{FN}$ | oLRP |
|---|---|---|---|---|---|---|---|
| WFD-1 | 0.8379 | 0.361 | 0.467 | 0.353 | 0.125 | 0.25 | 0.8 |
| WFD-2 | 0.776 | 0.346 | 0.47 | 0.341 | 0.174 | 0.296 | 0.806 |
| WFD-3 | 0.8327 | 0.398 | 0.452 | 0.322 | 0.13 | 0.286 | 0.77 |
| WFD-4 | 0.8338 | 0.382 | 0.511 | 0.3343 | 0.12 | 0.214 | 0.778 |
| WFD-5 | 0.8413 | 0.329 | 0.375 | 0.349 | 0.148 | 0.164 | 0.78 |
| WFD-C | **0.8639** | 0.33 | 0.357 | 0.349 | 0.158 | 0.143 | 0.77 |

The outputs in the Table 10 and Figure 19 obtained regarding the ensemble models developed within the scope of the study indicate that the highest AP50 score was 0.8639 in the WFD-C ensemble model. It is possible to state upon comparison of AP50 score of 0.754 in the PAA model, which was the highest result in the detection before the ensemble with the WFD-C model, that an increase of over 10% in AP50 score was achieved with the best ensemble model developed.

The bounding box outputs achieved from fracture detection performed with ensemble models in wrist X-ray images are provided in Figures 20 and 21 below as a sample for the right/left hand in the dataset.

WFD-1

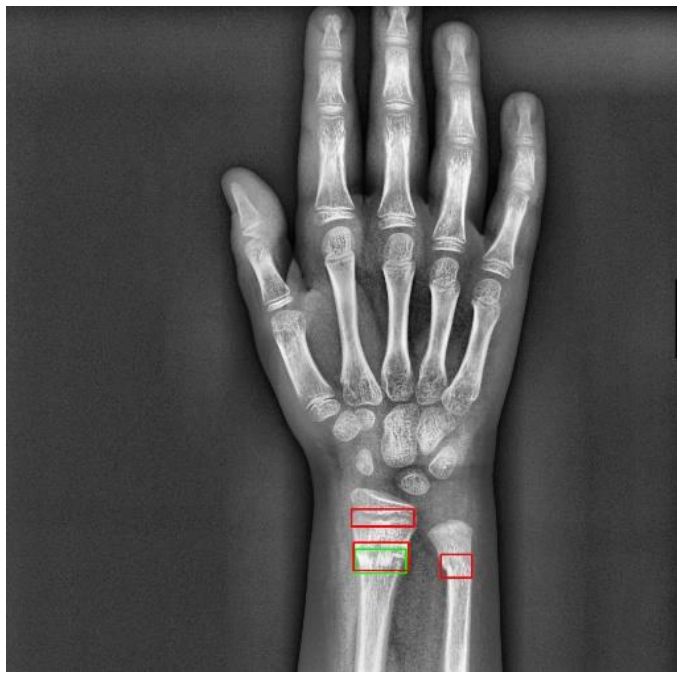

WFD-2

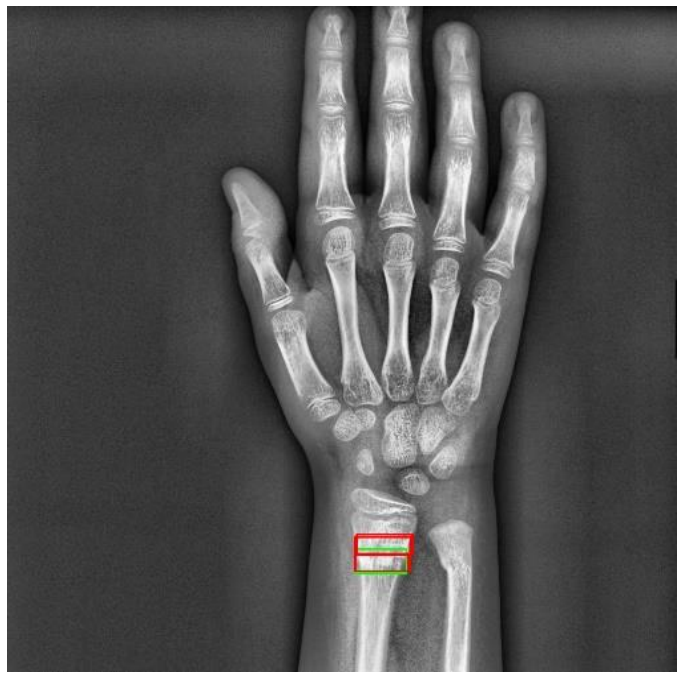

WFD-3

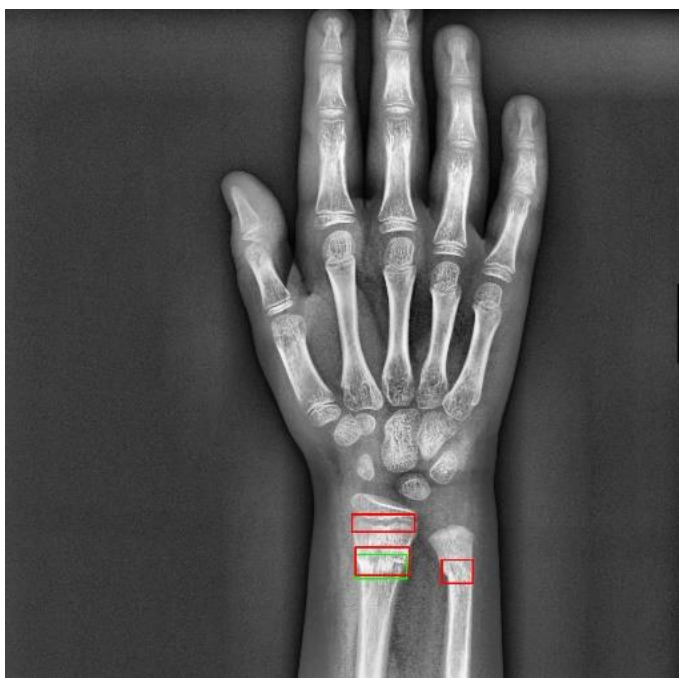

WFD-4

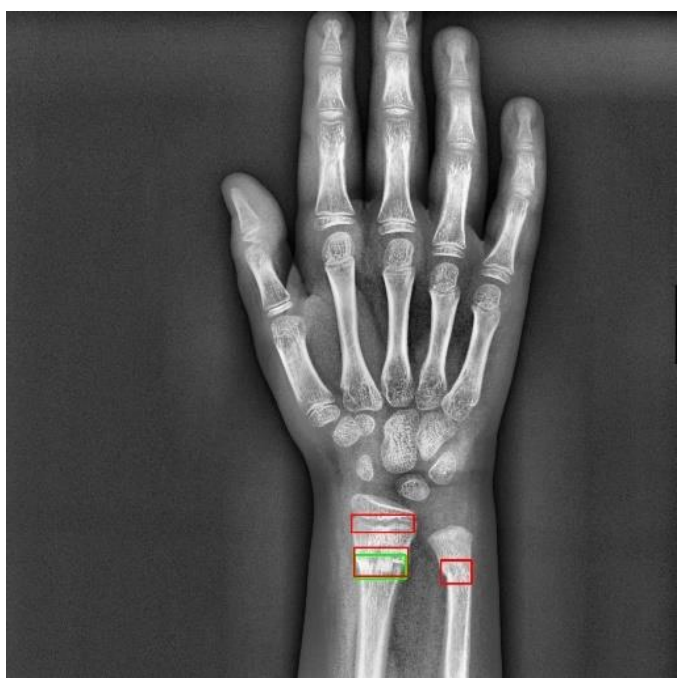

WFD-5

WFD-C

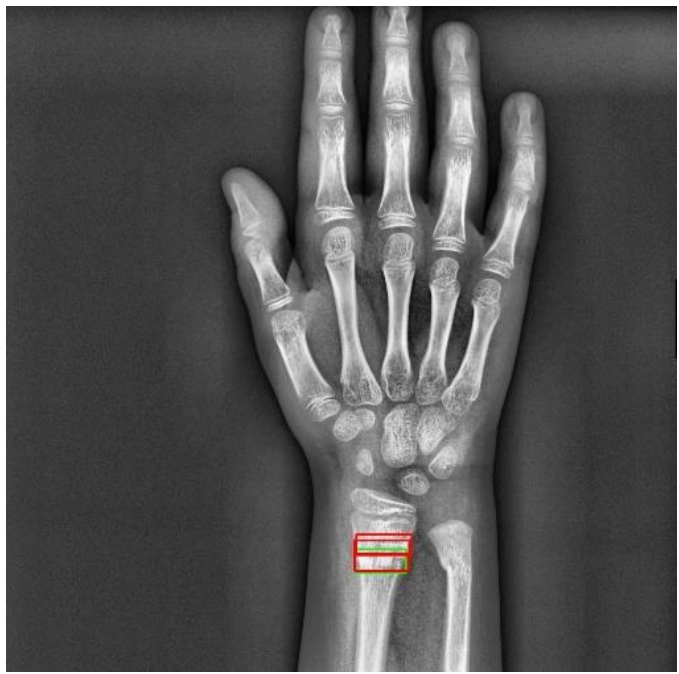
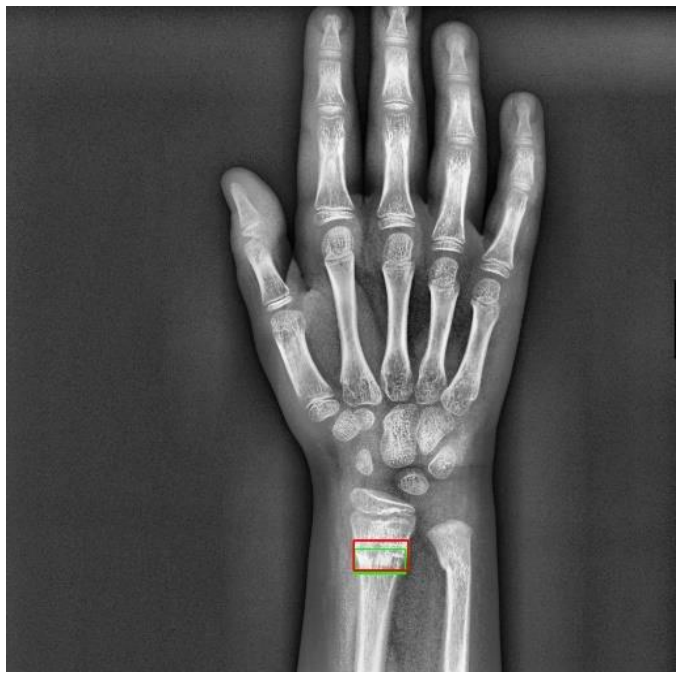

**Figure 20.** Sample of right wrist fracture results [ground-truth bounding box (green), predicted bounding box (red)].

WFD-1

WFD-2

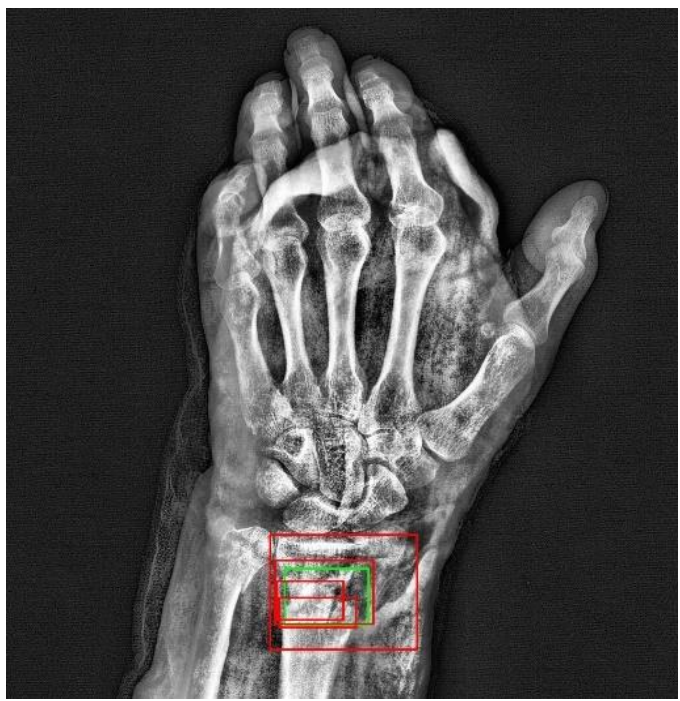
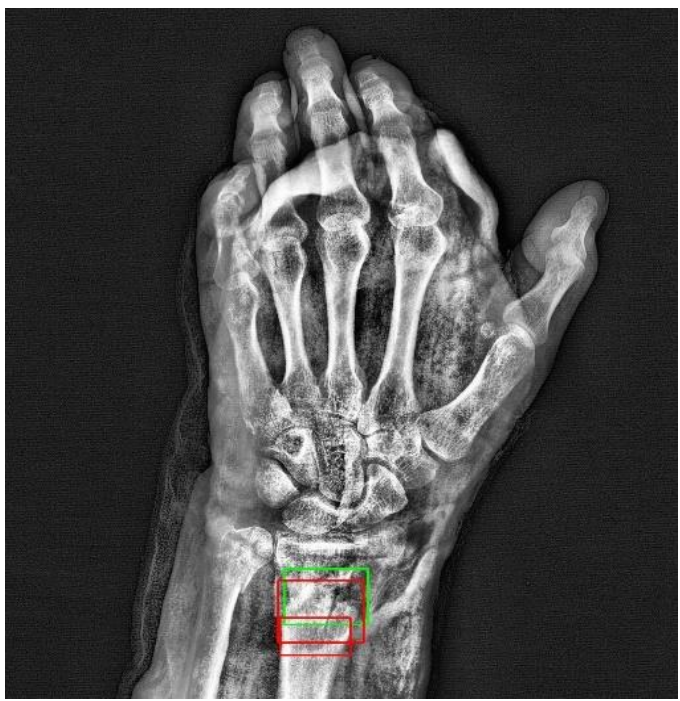

WFD-3

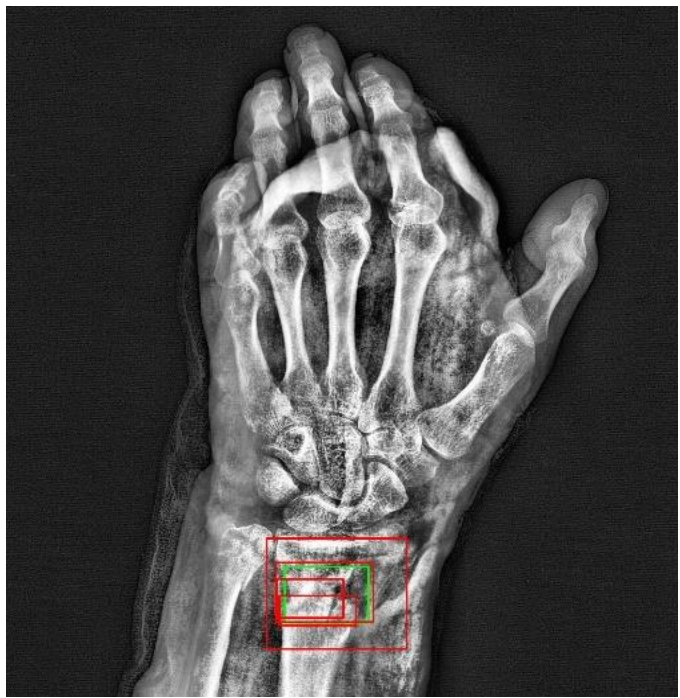

WFD-4

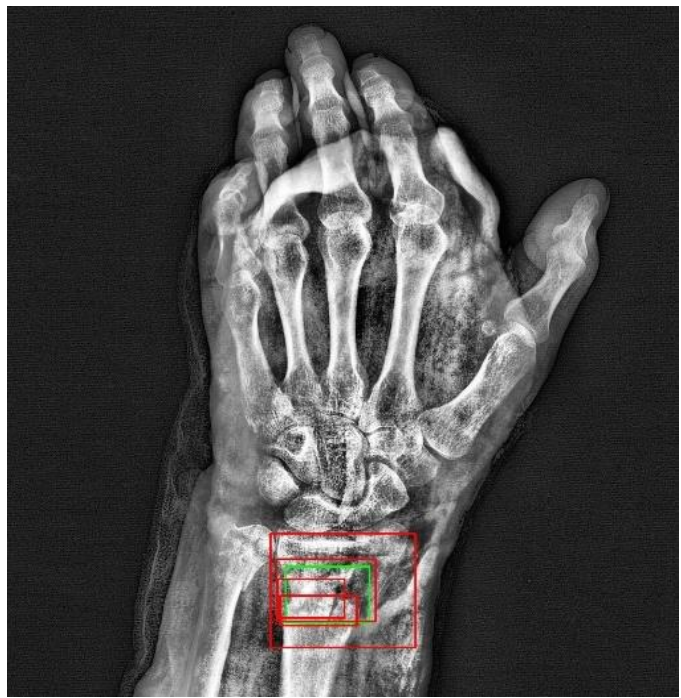

WFD-5

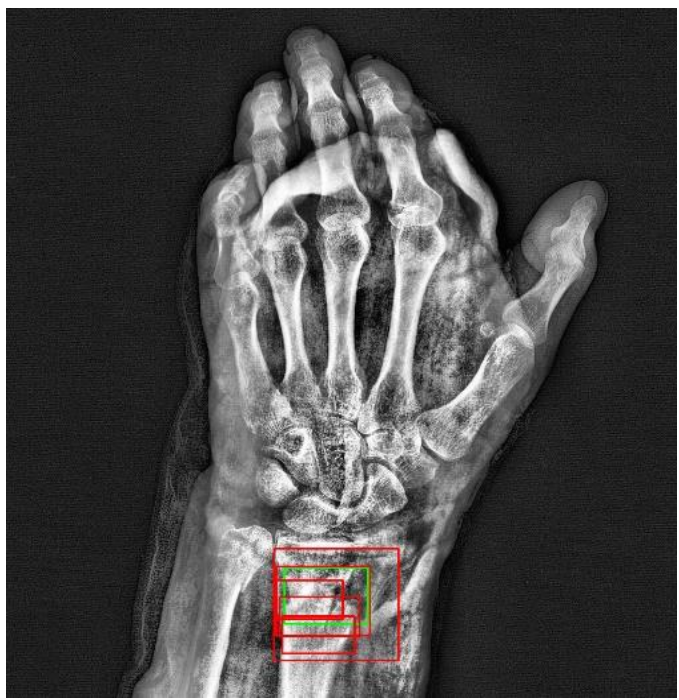

WFD-C

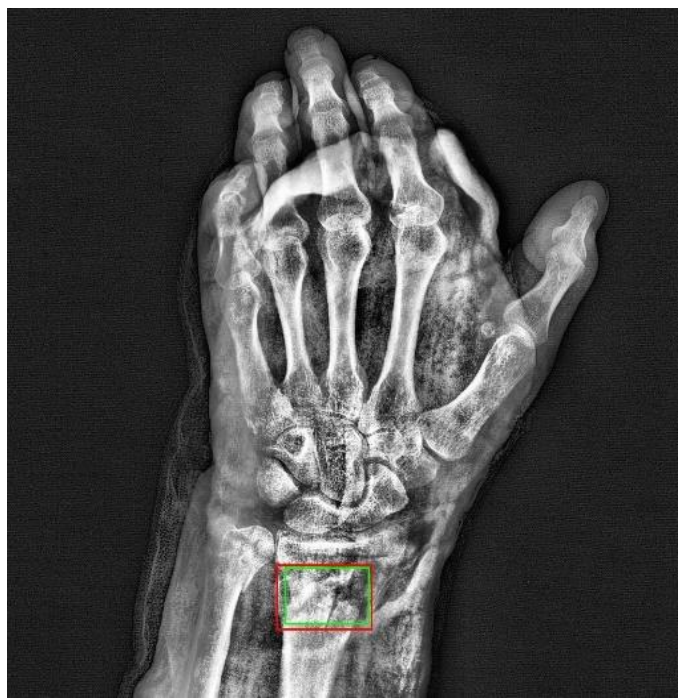

**Figure 21.** Sample of left wrist fracture results [ground-truth bounding box (green), predicted bounding box (red)].

The predicted bounding boxes achieved fracture detection suggest that, as shown in the images in Figures 20 and 21, the model that provides the most accurate results that are closest to the ground-truth bounding boxes is the WFD-C ensemble model developed in this study. Moreover, the examination of the number of predicted bounding boxes reveal that the model with the lowest number on the test data is also WFD-C. For the fracture detection in wrist X-ray images, Figure 22 below shows the total number of predicted bounding boxes obtained for each model on the test dataset as the result of the fracture detection performed with a total of 26 deep learning models, six of which are ensemble models.

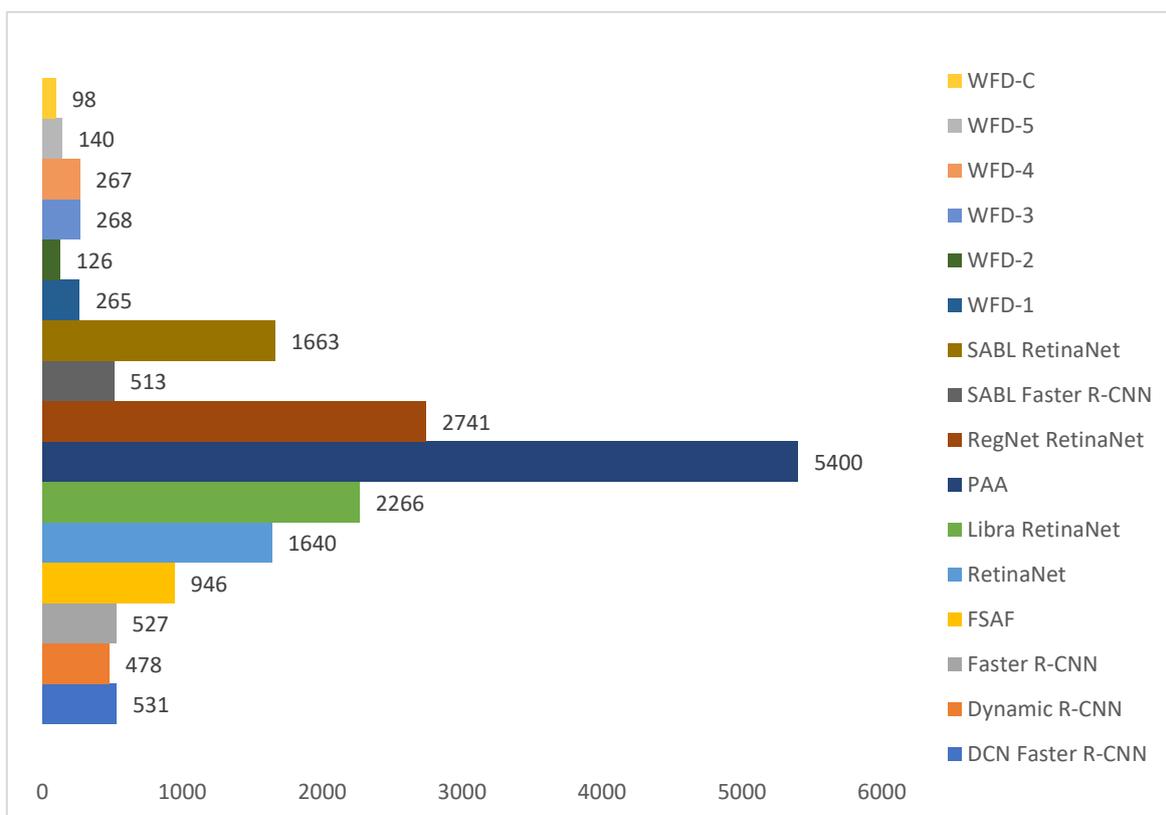

**Figure 22.** Count of predicted bounding box.

For the results of detection carried out with the ensemble models, the precision-recall curve of the WFD-C ensemble mode with the highest AP score is shown in Figure 23 below.

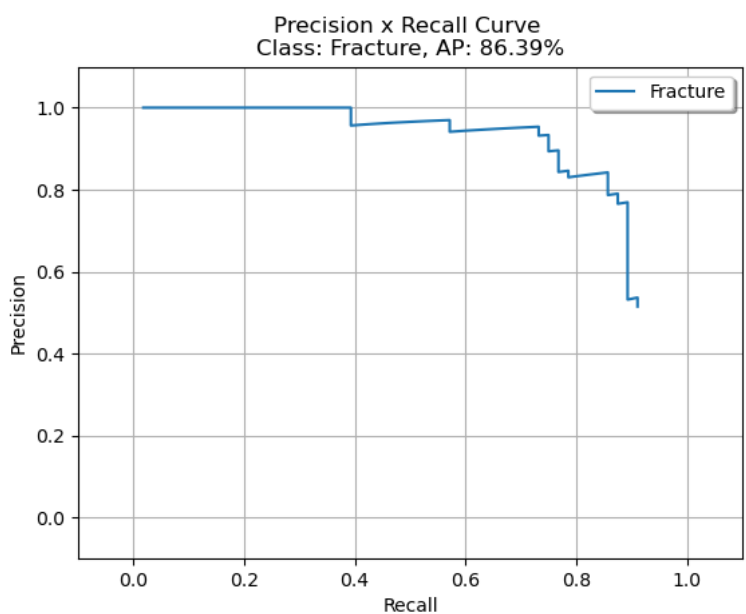

**Figure 23.** Precision-recall curve of WFD-C (best score of ensemble models).

When the results given in the table are examined and analyzed in terms of AP50, while the PAA model result was 0.754 in the Gazi University Hospital test dataset, the WFD-C model result was 0.8639, and an increase of 0.1099 was observed. In addition, when the test amount is doubled in the Gazi University Hospital dataset, it is seen that the developed WFD-C model achieves better AP results than the PAA model. When Table 11 is examined, it is understood that WFD-C has better detection results than the YOLO [42] model on a different number of datasets. When all these results are examined, it is understood that an increase in AP50 values was observed on a different number of datasets of the developed ensemble model and a contribution to the literature was obtained by obtaining good results.

**Table 11.** Comparison with various amounts of wrist test datasets.

| Model | Input | Dataset | Amount | AP50 | AR | LRP$_t$ | oLRP$_L$ |
|---|---|---|---|---|---|---|---|
| PAA | 800 × 800 × 3 | Gazi | 54 test | 0.754 | 0.496 | 0.56 | 0.310 |
| YOLOv3 | 750 × 750 × 3 | Gazi | 54 test | 0.531 | 0.298 | 0.164 | 0.378 |
| Proposed WFD-C | 800 × 800 × 3 | Gazi | 54 test | 0.8639 | 0.33 | 0.357 | 0.349 |
| PAA | 800 × 800 × 3 | Gazi | 54 test + 54 valid | 0.629 | 0.499 | 0.552 | 0.304 |
| YOLOv3 | 750 × 750 × 3 | Gazi | 54 test + 54 valid | 0.516 | 0.286 | 0.164 | 0.364 |
| Proposed WFD-C | 800 × 800 × 3 | Gazi | 54 test + 54 valid | 0.709 | 0.344 | 0.454 | 0.315 |

## 5. Conclusions and Future Work

Within the scope of this study, the aim was to develop the most compatible model for performing fracture detection in wrist X-ray images. The clinical dataset collected from Gazi University Hospital was used as the dataset in the study. Following the data preprocessing on the data, fracture detection was performed using 10 different object detection models based on deep learning. Subsequently, following experiments on various data augmentation methods and by performing augmentation on the training dataset using the method with the greatest contribution, new detection procedures were carried out with these 10 models. After 20 different procedures of fracture detection performed with the deep-learning-based models available in the literature, the results achieved from these procedures were examined from different perspectives, and six different ensemble models were developed to further improve the results of detection. As the result of the procedures of fracture detection performed using 26 different models in total, the highest AP score was obtained using the WFD-C ensemble model developed within the scope of this study. The contributions of the study to the literature are as follows.

- Because a clinical dataset was used as a dataset in the study, the new ensemble model developed within the scope of the study has the potential to be used in hospitals in the future, thus contributing to the literature.
- The augmentation method used in this study determined and used specifically for this study as the result of the data augmentation methods following the data preprocessing performed will be a good reference for those who will work on similar types of (medical X-ray) images in the future, constituting another contribution of the study.
- For the evaluation of the results of fracture detection, in addition to AP and AR scores, which are among the parameters currently available in the literature, LRP parameters were calculated on medical data for the first time in the literature within the scope of this study.
- To further improve the best result achieved in 20 different procedures of detection performed with deep-learning-based models available in the literature, the results were examined from five different perspectives to develop ensemble models. An approximately 10% increase in AP score was achieved with the best ensemble model developed based on weight box fusion to further improve the fracture detection results. With the new ensemble model developed with this approach, a unique detection model that will contribute to the literature was created.

Within the scope of this study, in which fracture detection was performed in wrist X-ray images, the aim is to provide assistance to physicians who are not specialized in their fields and/or especially to those working in emergency services in diagnosing fractures on X-ray images to allow them to apply the required treatments. Further in the study, an application can be developed to assist physicians, which can be used on portable devices such as mobile phones, tablets and laptops by operating in real-time, by studying other types of bone fractures that are frequently encountered in emergency services in addition to wrist images. In addition, if there is a portable X-ray in medical vehicles sent to help people in major disasters, epidemics or countries with underdeveloped health systems, evaluation can be made without the need for a radiologist. Regarding fracture detection in wrist bone X-ray images, in future works, in addition to fracture detection, an application can be developed that can perform classification, fracture detection and segmentation processes in normal and abnormal (fracture) image datasets, and that physicians can use on a portable device. For this purpose in the future, a hybrid system can be developed by using deep learning and various machine learning methods, especially for fracture detection processes. When the study is considered especially in terms of the dataset, information on inclusion/exclusion criteria is given in Table 12.

**Table 12.** Inclusion/exclusion criteria for dataset.

| Inclusion Criteria | Exclusion Criteria |
| --- | --- |
| Fracture labeling: Only fractures in the radius and ulna bones are labeled in Wrist. | Fracture labeling: other small bone (trapezoid, trapezium, scaphoid, capitate, hamate, triquetrum, pisiform, lunate) fractures in Wrist were not studied and ignored. |
| Image size: Images were rescaled to 800 × 800 × 3, and deep learning models supporting this size were used. | Image size: YOLO and other deep learning models that do not support 800 × 800 × 3 size were not used. |
| Data collection process: X-ray images of patients from the last 10 years (between 2010 and 2020) were used in 2020. | Data collection process: X-ray images of patients after 2020 were not used. |
| Number of fractures: there are 570 fractures in 542 images. Thus, there are multiple fracture ones in an image. | Number of fractures: images with not just one fracture, but one and/or more than one fracture were used. |
| The number of patients with fractures in both hands is 11. The distribution of these patients is 10 in Train and 1 in Test. | The patient with a fracture in both hands was not included in the validation. |
| The number of patients under the age of 12 and adults are 21 and 254, respectively. There is heterogeneity in these patient numbers. | There is no homogeneity, that is, an equal distribution, in the number of patients under the age of 12 and adults. |
| One radiologist and 2 orthopedists were jointly involved in the labeling of the fractures. | No more than 3 physicians were used in the labeling of fractures. There is no difference of opinion in the labeling of physicians. |
| In the Dataset distribution, the number of fractures per image is highest in the train dataset. | In the distribution of the dataset, the number of fractures per image was not considered to be equal in the train, validation and test dataset. |
| In the dataset, the number of females is 134 and the number of males is 141. | Equality in the number of males and females was not considered in the dataset. |
| The images in the dataset are 7% pair (right, left), 43% right-hand, 50% left hand. | Equality was not observed in the number of pairs, right-handed and left-handed images in the dataset. |
| Since the graphics card in the local PC hardware used in the study is 4 GB, object detection models that support this are used in MMDetection. | In the study, models that support a graphics card higher than 4 GB from the object detection model-hands in MMDetection could not be used. |

**Author Contributions:** Conceptualization, F.H., F.U. and O.P.; methodology, F.H., F.U. and O.P.; software, F.H., F.U. and O.P.; validation, F.H., F.U. and O.P.; formal analysis, F.H., F.U. and O.P.; investigation, F.H., F.U. and O.P.; resources, F.H., F.U. and O.P.; data curation, F.H., F.U. and O.P.; writing—original draft preparation, F.H., F.U. and O.P.; writing—review and editing, F.H., F.U., O.P., M.Ç., T.T., N.T., U.K., B.D. and F.M.; visualization, F.H., F.U. and O.P.; supervision, F.H., F.U. and O.P. All authors have read and agreed to the published version of the manuscript.

**Funding:** This research received no external funding.

**Institutional Review Board Statement:** The study was conducted according to the guidelines of the Declaration of Helsinki, and approved by the Ethics Committee of Gazi University (27.04.2020-E.51150).

**Informed Consent Statement:** Not applicable.

**Data Availability Statement:** Not applicable.

**Conflicts of Interest:** The authors declare no conflict of interest.

Code is available at https://github.com/fatihuysal88/wrist-d

Personal website of corresponding author: https://fatihuysal88.github.io